\documentclass[usenatbib]{mn2e}
\usepackage[pdftex]{color}
\usepackage[draft]{graphicx}
 \usepackage[draft]{hyperref}
\usepackage[fleqn]{amsmath}     
\usepackage{amssymb}

\usepackage{longtable}
\usepackage{lscape}

\title[Population-Orbit Superposition: Method Validation]{Disentangling the formation history of galaxies via population-orbit superposition: method validation}
\author[Zhu et al]{Ling Zhu$^1$\thanks{Corr author: lzhu@shao.ac.cn}, Glenn van de Ven$^2$, Ryan Leaman$^3$, Robert J. J. Grand$^4$, \and Jes\'us Falc\'on-Barroso$^{5,6}$, Prashin Jethwa$^2$, Laura L. Watkins$^7$, Shude Mao$^8$, \and Adriano Poci$^{9,10}$, Richard M. McDermid$^9$, Dandan Xu$^8$, Dylan Nelson$^4$ \\
  $^1$ Shanghai Astronomical Observatory, Chinese Academy of Sciences, 80 Nandan Road, Shanghai 200030, China\\
  $^2$ Department of Astrophysics, University of Vienna, T\"urkenschanzstrasse 17, 1180 Wien, Austria\\
  $^3$ Max Planck Institute for Astronomy, K\"onigstuhl 17, 69117 Heidelberg, Germany \\
  $^4$ Max-Planck-Institut f\"{u}r Astrophysik, Karl-Schwarzschild-Str. 1, 85748 Garching, Germany\\
  $^5$ Instituto de Astrofisica de Canarias C/ Via Lactea, s/n, E-38205, La Laguna, Tenerife, Spain\\
  $^6$ Departamento de Astrof\'isica, Universidad de La Laguna, E-38200 La Laguna, Tenerife, Spain\\
  $^7$ AURA for ESA, Space Telescope Science Institute, 3700 San Martin Drive, Baltimore MD 21218, USA \\
  $^8$ Physics Department and Tsinghua Centre for Astrophysics,
  Tsinghua University, Beijing 100084, China\\
  $^9$ Astronomy, Astrophysics, and Astrophotonics Research Centre, Department of Physics and Astronomy, Macquarie\\ University, NSW 2109, Australia\\
  $^{10}$ European Southern Observatory, Karl-Schwarzschild-Str. 2, 85748 Garching bei M\"unchen, Germany\\
}

\begin{document}
\date{}
\maketitle

\begin{abstract}
We present population-orbit superposition models for external galaxies based on Schwarzschild's orbit-superposition method, by tagging the orbits with age and metallicity. The models fit the density distributions, as well as kinematic, age and metallicity maps from Integral Field Unit (IFU) spectroscopy observations. We validate the method and demonstrate its power by applying it to mock data, similar to those obtained by the Multi-Unit Spectroscopic Explorer (MUSE) IFU on the Very Large Telescope (VLT). These mock data are created from Auriga galaxy simulations, viewed at three different inclination angles ($\vartheta=40^o, 60^o, 80^o$). Constrained by MUSE-like mock data, our model can recover the galaxy's stellar orbit distribution projected in orbital circularity $\lambda_z$ vs. radius $r$, the intrinsic stellar population distribution in age $t$ vs. metallicity $Z$, and the correlation between orbits' circularity $\lambda_z$ and stellar age $t$. A physically motivated age-metallicity relation improves recovering the intrinsic stellar population distributions. We decompose galaxies into cold, warm and hot + counter-rotating components based on their orbit circularity distribution, and find that the surface density, mean velocity, velocity dispersion, age and metallicity maps of each component from our models well reproduce those from simulation, especially for projections close to edge-on. These galaxies exhibit strong global age vs. $\sigma_z$ relation, which is well recovered by our model. The method has the power to reveal the detailed build-up of stellar structures in galaxies, and offers a complement to local resolved, and high-redshift studies of galaxy evolution.
\end{abstract}


\section{Introduction}
\label{S:intro}

Stellar dynamics provides a fossil record of the formation history of galaxies. Stars that were born and remain in quiescent environments tend to be on regular rotation-dominated orbits. On the other hand, stars born from turbulent gas or that have been dynamically heated after birth will be on warmer orbits with more random motions \citep{Leaman2017}. Stellar heating mechanisms include violent mergers \citep[e.g.][]{Benson2004,House2011, Helmi2012, Few2012, Ruiz-Lara2016} and long-term secular heating of the disk via internal instabilities \citep[e.g.][]{Jenkins1990, Aumer2016, Grand2016}. 

As the Universe evolves, a galaxy's mass density, gas fraction and star formation all decrease. This likely reduces the velocity dispersion of the gas from which stars form, the mass spectrum of dense giant molecular clouds, and the frequency of mergers \citep{Genzel2011, Wisnioski2015}. Stellar kinematics are therefore expected for multiple reasons to be systematically correlated with stellar ages \citep{Trayford2018}. Several observations have revealed that old stars dominate the light of random-motion-dominated bulges, while younger stars live on thinner disks. Stars born during the same epoch tend to live on similar orbits (e.g., \citealt{Bird2013}, \citealt{Stinson2013}). At the present day, the stellar phase space distribution of a galaxy is thus a combination of stars formed over its lifetime. A mixture of merging events and star formation episodes determines the diversity of a galaxy's structure and stellar populations. 

Observationally, it is difficult to identify coherent structures in the density distribution and kinematics of stars formed at high redshift \citep{vdW2016}.
Fortunately, the chemistry and age imprinted in a star provide coordinates of the time and environment of its birth. Taking the Milky Way (MW) as an example, the chemical abundance as well as the 6D phase-space information of a single star could be obtained by combining {\em Gaia} \citep{Gaiadr2} and spectroscopic surveys. In the solar neighborhood, most stars are disk stars, which we can identify as they are on near-circular orbits, and are both young and metal-rich \citep[e.g.][]{Mackereth2017}. Although spatially-coincident, we can also identify a small fraction of halo stars as they are on radial/vertical-motion-dominated orbits, and are old and metal-poor \citep[e.g.][]{Helmi2018, Belokurov2018, Belokurov2019}. The chemical information of disk stars and halo stars gives us insight into the formation history of the MW \citep{Helmi2018, Belokurov2018, Fattahi2019}.

However, only a handful of galaxies are near enough for us to resolve their stars. For most external galaxies, all of our information comes from integrated light. In these cases, the spectrum we observe at each pixel is a light-weighted combination of spectra from all the stars along the line-of-sight, which come from different populations with different ages, metallicities and kinematics. By full spectrum fitting (e.g. \citealt{Cappellari2017}), we can obtain the line-of-sight velocity distribution (LOSVD), which is usually described by a Gauss-Hermite (GH) profile with parameters of mean velocity ($V$), velocity dispersion ($\sigma$), and/or higher order GH coefficients, like the third and fourth order $h_3$ and $h_4$ or even higher $h_5$ and $h_6$. These full spectrum fits also return the average age and metallicity of the underlying stellar populations. Such methodologies have been applied in many Integral Field Unit (IFU) spectroscopic surveys, such as the Calar Alto Legacy Integral Field Area Survey (CALIFA; \citealt{Sanchez2012}), the Sydney AAO Multi-object Integral Field galaxy survey (SAMI; \citealt{Croom2012}), and the Mapping Nearby Galaxies at APO survey (MaNGA; \citealt{Bundy2015}). These surveys provide a spectrum at each pixel across the galaxy plane. From these spectra and the aforementioned techniques, we obtain kinematic maps ($V$, $\sigma$, $h_3$, $h_4$...), as well as age and metallicity maps.

Disentangling the different stellar populations in present-day galaxies, and the structures they form, will give us insight into the galaxy's formation history; however this is challenging as it typically requires resolved stellar abundances and ages or deep integrated light spectroscopy \citep{Leaman2013, Boecker2019}. Full spectrum (or SED) fitting has been pushed to provide not only average stellar population properties, but a distribution of ages and metallicities - such as a star formation history (SFH) or age metallicity relation (AMR) \citep[e.g.][]{Cid2005, Cappellari2017, Carnall2019, Leja2019, McDermid2015}. Based on the SFH obtained at each pixel, galaxies can be decomposed into structures with different stellar ages and metallicities (\citealt{Guerou2016}, \citealt{Pinna2019a}, \citealt{Pinna2019b}, \citealt{Pizzella2018}, \citealt{Tabor2019}).

Dynamical models offer us an alternative and powerful tool to probe a galaxy's formation history. The particle-based Made-to-Measure method \citep[M2M][]{deLorenzi2007, Long2010, Hunt2014} and the orbit-based Schwarzschild method \citep{vanderMarel&Franx1993, Rix1997, Cretton1999, Gebhardt2000, Valluri2004, vdB2008, vdV2008, Vasiliev2019} probe how stars orbit in a gravitational potential without ad-hoc assumptions about the underlying orbital structures. The triaxial Schwarzschild model developed by \citet{vdB2008} has proved to be effective at recovering the orbit distributions of a variety of galaxies (\citealt{Zhu2018b}; \citealt{Zhu2018a}; \citealt{Jin2019}). It has notably been applied to a large sample of 300 CALIFA galaxies in the local universe to recover their stellar orbit distributions \citep{Zhu2018a}. However, the orbits recovered in that study (and most others) are monochromatic and provide no information about the underlying stellar populations.

Recently, there have been a few pioneering works that have moved beyond this monochromatic view by tagging particles or orbits in dynamical models with a characteristic chemistry or age indicator. These works include a chemodynamic M2M model of the MW bulge \citep{Portail2017}, and both M2M \citep{Long2016} and Schwarzschild \citep{Long2018} chemodynamic models of four nearby galaxies. However, the power and limitations of these methods have not been characterised by testing against mock data. This is what we set out to do here.

Starting from the Schwarzschild code developed by \citet{vdB2008}, we arrive at a new population-orbit superposition method. Under the assumption that stars on the same orbit were born close in space and time, we tag each orbit in the Schwarzschild model with an age and metallicity. Thus if we imagine observing the model as we would in a real galaxy, we can predict not only the kinematic distribution along the line of sight but the age and metallicity properties as well. In this way, stellar populations at different positions are connected {\it by the underlying orbits}, providing a holistic model of the galaxy. A rather similar approach was recently applied to an edge-on galaxy, NGC 3115 \citep{Poci2019}, which offers a tantalizing view into the power of the method by providing the global stellar age vs. dispersion ($\sigma_z$) relation in an external galaxy.

In this paper, we validate our population-orbit superposition method and demonstrate its power in recovering dynamical structures of different stellar populations by using MUSE-like mock data created with a range of projections from a variety of simulated galaxies. The paper is organized in the following way: in Section~\ref{S:data}, we describe the mock data created from the simulations; in Section~\ref{S:method}, we describe the method; in Section~\ref{S:result}, we illustrate the model recovery of intrinsic orbit distribution, stellar population distribution, and the correlation in between for each galaxy; and in Section~\ref{S:decomp}, we illustrate the orbital decomposition of the galaxies and show the recovery of the age and metallicity properties of different components. We discuss the results in Section~\ref{S:dis}, and summarize in Section~\ref{S:summary}.

\section{Mock data}
\label{S:data}
\subsection{Simulations}
\label{SS:simulation}
The simulations used for our study are taken from the Auriga project \citep{Grand2017, Grand2019}, which is a suite of 40 cosmological magneto-hydrodynamical simulations for the formation of the Milky Way-mass haloes. These simulations were performed with the AREPO moving-mesh code \citep{Springel2010}, and follow many important galaxy formation processes such as star formation, a model for the ionising UV background radiation, a model for the multi-phase interstellar medium, mass loss and metal enrichment from stellar evolutionary processes, energetic supernovae and AGN feedback and magnetic fields \citep{Pakmor2017}. We refer the reader to \citet{Grand2017} for more details. In this study, we select three galaxies from the Auriga simulation suite at a mass resolution of $\sim 5\times10^4\, M_{\odot}$ for baryons. The comoving gravitational softening length for the star particles and high-resolution dark matter particles is set to 500 $h^{-1}$ cpc. The physical gravitational softening length grows with the scalefactor until a maximum physical softening length of 369 pc is reached. This corresponds to z = 1, after which the softening is kept constant. The details of which are listed in Table~\ref{tab:simulation}.

\begin{table*}
\caption{The 3 simulated galaxies from Auriga project. From left to right: the galaxy name, stellar mass $M_\star$, neutral hydrogen mass $M_{\mathrm{HI}}$ \citep{Marinacci2017}, dark matter (DM) halo mass $M_{\rm 200}$, stellar particle resolution $M_{\rm StarParticle}$, DM particle resolution $M_{\rm DMParticle}$ in unit of solar mass $M_{\odot}$, Hubble types and specific properties, the inclination angles ($\vartheta$ in degree) projected with for creating mock data sets. }
\scriptsize\centering
\label{tab:simulation}
\begin{tabular}{*{10}{l}}
\hline
\hline
Name  & $M_\star$ & $M_{\mathrm{HI}}$ & $M_{\rm 200}$
  & $M_{\rm StarParticle}$ & $M_{\rm DMParticle}$ &  type  &  $\vartheta(^\circ)$  \\
\hline
$Au-5$  &$ 6.7e10$   & $7.2e9 $ &  $1.2e12$ & $\sim 5e4$ &  $\sim 3e5$ &Spiral: spiral arms, weak bar  & $40,60,80$ \\
$Au-6$  &$ 4.75e10$   & $1.5e10$ &$1.0e12$ &  $\sim 5e4$ & $\sim 3e5$ &Spiral: spiral arms, weak bar  & $40,60,80$ \\
$Au-23$  &$ 9.02e10$ & $1.45e10$ &$1.6e12$& $\sim 5e4$ & $\sim 3e5$  &Spiral: warps, strong bar  & $40,60,80$ \\
  \hline
 \end{tabular}
\end{table*}

\subsection{Mock data}
\label{SS:kindata}

From each simulation, we take three projections with inclination angles of $\vartheta = 80^o,\, 60^o\, 40^o$ (from edge-on to face-on). Then, we create a mock dataset for each projection, thus we have $3 \times 3 = 9$ mock galaxies in total. Au-6 $\vartheta = 80^o$ is taken to illustrate the method throughout the paper.

The mock data are created as follows. We first project a simulation to the observational plane with inclination angle $\vartheta$ ($80^o, 60^o, 40^o$), and place it at distance $d = 30$ Mpc, then observe it with pixel size of 1 arcsec ($1'' = 145 \, {\rm pc}$). Then we calculate the stellar mass of particles in each pixel to obtain a surface mass density map. According to the number of particles in each pixel, we then perform a Voronoi binning process to reach a signal-to-noise ratio of $S/N=50$, assuming Possion noise $\sim \sqrt{N_{\rm particles}}$. With all the particles in each Voronoi bin, we obtain the mass-weighted mean velocity, velocity dispersion, $h_3$, $h_4$ by fitting a GH profile (\citealt{Gerhard1993}; \citealt{vanderMarel&Franx1993}) to the stellar LOSVD, and calculate mass-weighted average age ($t$) and metallicity ($Z/Z_{\rm sun}$). 

After the voronoi binning, the spatial resolution of our mock data is $\sim 150-1000$ pc. Considering the softening length of 369 pc for star particles in the simulation, the actually spatial resolution is $\sim 400-1000$ pc, which is comparable, but slightly lower than that of the kinematic data (binned with $S/N=100$) from the Fornax 3D project \citep{Sarzi2018}. 

We use a simple function inferred from the CALIFA data to construct the errors for kinematic maps \citep{Tsatsi2015}, but here considering higher $S/N$. For age and metallicity, the observational errors are more complicated. Tests on full-spectrum fitting to mock spectra of $S/N=40$ obtained random errors of $10\%$ for age and metallicity \citep{Pinna2019a}, the errors could be lower for spectra with higher $S/N$, while it could be higher for real spectra due to possible systematic effects. For this proof-of-concept we adopt relative errors of $10\%$ for age and metallicity. The kinematics, age and metallicity maps are then perturbed by random numbers, normally distributed with dispersions equal to the observational errors. 
The error maps of the mock data are similar to the data of MUSE observations for galaxies in the Fornax 3D project \citep{Sarzi2018}.

\begin{figure*}
\centering\includegraphics[width=10cm]{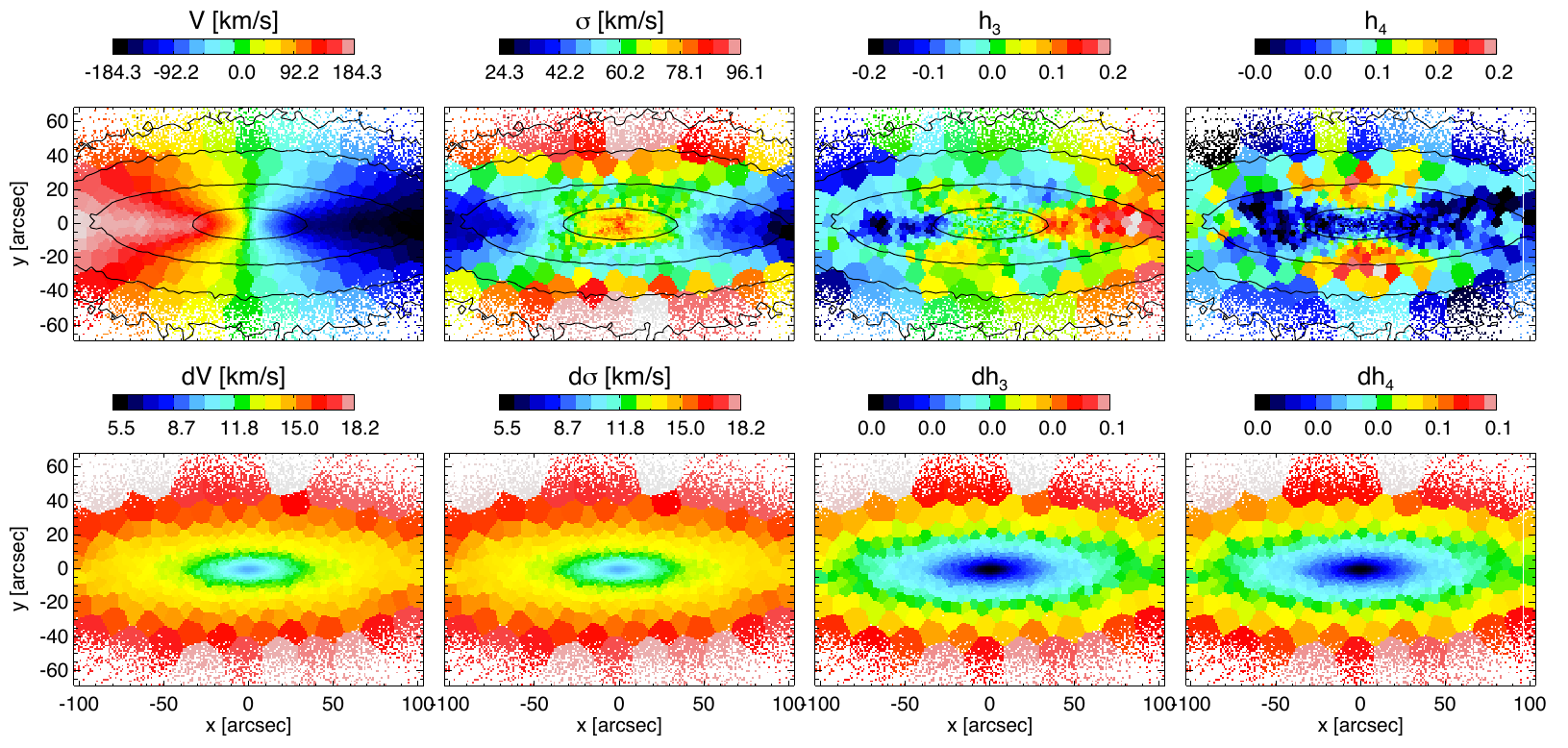}\includegraphics[width=5.7cm]{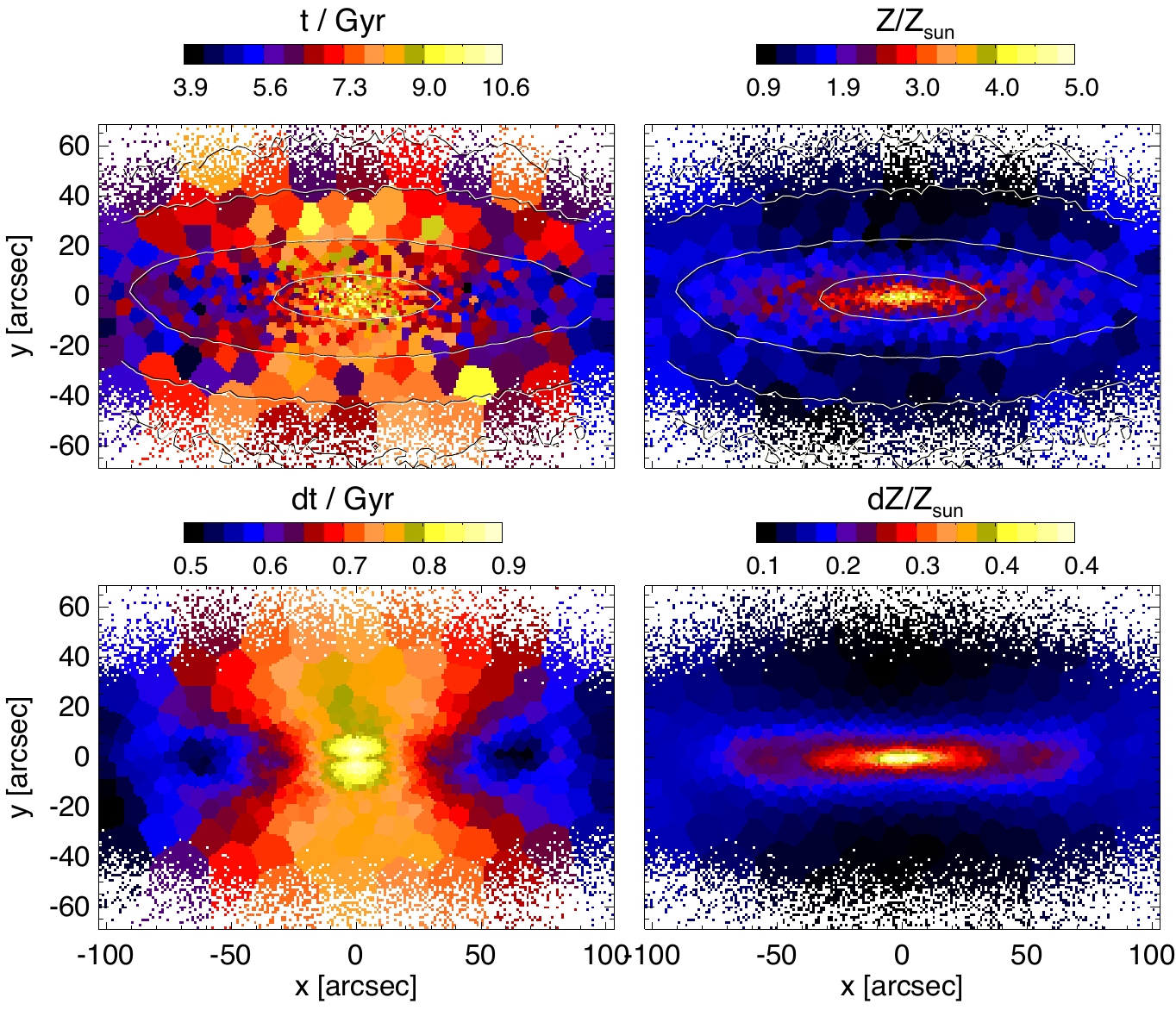}
\caption{MUSE-like mock data created from the simulation
  Au-6, projected with inclination angle of $\vartheta =
  80^o$. The six columns from left to right are mean velocity $V$,
  velocity dispersion $\sigma$, GH coefficients $h_3$ and
  $h_4$, age and metallicity. The first row are perturbed data, the
  second row are errors ($dt$ and $dZ$ are $10\%$ of the original unperturbed data). The overplotted contours represent the
  surface mass density. }
\label{fig:mockdata_muse}
\end{figure*}

The mock data created from the simulation Au-6 with $\vartheta = 80^o$ are shown in Figure~\ref{fig:mockdata_muse}. From left to right, they are stellar mean velocity $V$, velocity dispersion $\sigma$, GH coefficient $h_3$, $h_4$, age, and metallicity maps. The first row are the perturbed data and the second row are the corresponding error maps. 

For real galaxies, the kinematic maps obtained from observations are usually light-weighted,
In that case, we typically measure light-weighted age and metallicity
maps, and use surface {\em brightness} as the tracer density for
consistency. Orbits in the model should be
interpreted as light-weighted. 
Here, however, the mock kinematics, age, and metallicity maps are mass-weighted so that we use surface {\em mass} density - rather than surface
{\em brightness} - as the tracer density distribution. Therefore, the orbits in the model are mass-weighted.
For method validation, mass-weighted or light-weighted data do not make any difference. 

\section{Method}
\label{S:method}
In this section we describe how we fit the stellar kinematic maps as well as the
age and metallicity maps with a population-orbit superposition method.
The model will proceed as a two-step process: first, fitting the
kinematics maps with a standard Schwarzschild's orbit-superposition
model to obtain the orbit weights;
second, tagging the orbits with age and metallicities and fitting the age
and metallicity maps, to obtain the best-fit age and metallicity of the orbits.

\subsection{Schwarzschild method}
\label{SS:schw}
The three main steps to build a Schwarzschild model are: 1) create a
suitable model for the underlying gravitational potential; 2)
calculate a representative library of orbits within the
gravitational potential; and 3) find the combination of
orbits (solve the orbit weights) that match the observed kinematic
maps and luminosity/mass
distribution of the tracers.

The gravitational potential is constructed by a combination of stellar
mass distribution and dark matter halo. We de-project the surface brightness to 3D
luminosity density by assuming a set of viewing angles ($\vartheta,
\psi, \phi$). By multiplying the surface brightness by a stellar mass-to-light
ratio, we obtain the intrinsic stellar mass density.
Here for the mock galaxies, we actually use surface mass density,
instead of surface brightness, to construct the gravitational
potential. We still allow for a scale parameter $\alpha_{\rm star}$, which
is analogous to a mass-to-light
ratio, but with a true value of $1$, to be a free parameter.

A Multi-Gaussian Expansion (MGE) is
used for modelling the surface density and de-projection to 3D
density for the stellar component \citep{Emsellem1994, Cappellari2002}. We use the parameters describing intrinsic shapes ($q =
Z/X$, $p = Y/X$,
$u = \sigma_{\rm obs}/\sigma_{\rm intr}$) of the Gaussians, instead of the three viewing angles as free
parameters. X, Y, Z are the intrinsic major, intermediate and long axis of
the galaxy, $u$ is the ratio between $\sigma$ of observed long axis to
the intrinsic long axis. For galaxies close to axisymmetric, we fix $u
= 0.9999$, while $p$ and $q$ are left free, thus triaxiality of the
stellar component is still allowed.
The dark matter
is assumed to be a spherical NFW halo, with concentration $C$ fixed
according to $M_{\rm 200}$ vs. $C$  correlation of \citet{Dutton2014}.

In summary, we have four free parameters describing the gravitational potential:
the scale parameter of stellar mass $\alpha_{\rm star}$ (comparable to a stellar mass-to-light ratio), intrinsic shape
parameters $p$ and $q$, and dark matter virial mass $M_{\rm 200}$.

The method of orbit library sampling and model fitting follow exactly as
described in \citet{Zhu2018a} and \citet{vdB2008}, which we do not repeat
here. It should be emphasized that we do not fit $V$,
$\sigma$ maps directly, but rather the LOSVD expanded in GH coefficients $h_1$, $h_2$, $h_3$ and $h_4$ to solve the
orbit weights. However, we extract $V$ and $\sigma$ maps from the model at the end for
direct comparison to the observational data.
By exploring the free parameters describing the gravitational
potential, we find the best-fitting model which reproduces the
observed stellar kinematic maps and mass distribution.

\begin{figure}
\centering\includegraphics[width=8cm]{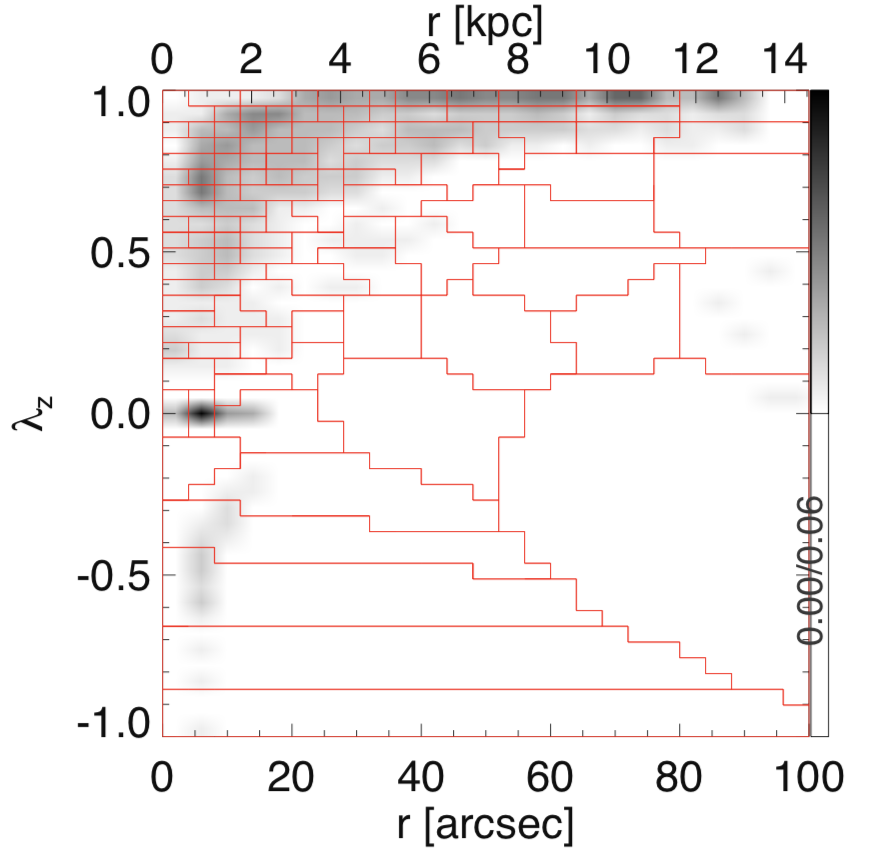}
\caption{Probability density of stellar orbits in the phase space of circularity
  $\lambda_z$ versus time-averaged orbital radius $r$. Darker color indicates higher probability as
  indicated by the colorbar, and the total orbit weight has been
  normalized to unity. This galaxy has a large fraction of highly
  circular tube orbits with $\lambda_z \sim 1$ extending to large
  radius. In the inner regions, there are more tube orbits with
  significant random motions with $\lambda_z \sim 0.5$, some radial-motion
  dominated box orbits $\lambda_z \sim 0$ , and
  a small fraction of counter-rotating orbits with $\lambda_z < 0$.
  The red polygons indicate the Voronoi binning scheme we have adopted in
  the phase-space, yielding $\sim 100$ orbit bundles,
  each with a minimum weight of 0.005.}
\label{fig:lzr_bin}
\end{figure}

We characterise the orbits with two parameters: time-averaged radius $r$
and circularity $\lambda_z$. Following \citet{Zhu2018a}, $\lambda_z$ is defined as the
angular momentum $L_z$ normalized by the maximum that
is allowed by a circular orbit with the same binding energy. 
All quantities are taken as
average of the particles sampled along the orbit over equal time interval.
The stellar orbit distribution of a galaxy is described as  
the probability density distribution in the phase-space of
$\lambda_z$ vs. $r$. Figure~\ref{fig:lzr_bin} illustrates the orbit
distribution of a typical spiral galaxy. Darker color indicates higher
probability; the total weight of the orbits has been normalized to unity.

The orbit library consists of a few thousand orbits, and a few hundred of them gain significant
weights at the end.
To reduce the noise in fitting age and metallicity
maps, we perform a Voronoi binning in the phase-space $r$
vs. $\lambda_z$, and divide the
orbits into $N_{\rm bundle}\sim 100$ bundles. Orbits with similar $r$
and $\lambda_z$ are included in the same bundle, to ensure each bundle has a minimum of orbit weight of
0.005. The resulting binning scheme is shown as the red polygons in
Figure~\ref{fig:lzr_bin}.

\subsection{Tagging stellar orbits with stellar populations}
\label{SS:tag}

The observed age map presents values of age, $t_{\rm obs}^i$, at
each aperture $i$ on the observational plane, with a total number of
$N_{\rm obs}$ apertures.  Throughout the paper, one aperture
indicates one spatial bin on the observational plane which may include a few pixels.
After dividing the orbits into $N_{\rm bundle}$ bundles (Figure~\ref{fig:lzr_bin}), we re-sample
particles from these orbits in each bundle, with the number of
particles proportional to their orbit weights.
Then we add up all the particles sampled from each orbit bundle,
project them to the observational plane, and calculate the mass $f_k^i
$ (mass for mass-weighted and flux for light-weighted models) contributed by the orbital bundle $k$ at each aperture $i$.

This orbit bundle k, is tagged with a single value of age $t_k$.
The average value of age in
each aperture $i$ is a linear average of the $N_{\rm bundle}$ orbital bundles:
\begin{equation}
  \label{eqn:agefit}
    t_{\rm obs}^i = \Sigma_{k=1}^{N_{bundle}} t_k f_k^i / \Sigma_k f_k^i,
\end{equation}
for $i = 1,...N_{\rm obs}$.
Similarly, for metallicity:
\begin{equation}
  \label{eqn:Zfit}
    Z_{\rm obs}^i = \Sigma_{k=1}^{N_{bundle}} Z_k f_k^i / \Sigma_k f_k^i.
\end{equation}
We then solve for the values of $t_k$ and $Z_k$ using a Bayesian statistical analysis, which we will describe in detail in Section~\ref{SS:Bayes}. 
 As we will see, reproducing on-sky age and metallicity maps may be
 possible, however to reproduce them with the correctly correlated
 combinations of age and especially metallicity is non-trivial.
 
\subsection{Age-metallicity correlation}
\label{SS:age-met}

We wish to adopt the most agnostic parameterization of the possible
metallicity and age values for each orbit bundle. Unfortunately, a
completely unconstrained age-metallicity parameter space results in
poor recovery of the known 2D distribution on age vs. metallicity (see in section~\ref{SS:AM_dis}).

In order to provide a theoretically motivated link between age and
metallicity which is flexible enough and unbiased for our purposes, we leverage the
statistical chemical scaling relations presented in \citet{Leaman2012}, and
model described by \citet{Oey2000}.  These essentially map a galaxy's
chemical evolution into a parameter space that is: 1) self-similar
across time and spatial scales for galaxies of different masses, and 2) is
easily expressed in a robust statistical functional form (binomial).

The shape of galaxy age-metallicity relations and metallicity
distribution functions show mass dependent behaviours \citep[e.g.][]{Kirby2013, Leaman2013}. However \citet{Leaman2012} identified that in
linear metal fraction ($Z/Z_{\odot}$), all Local Group galaxies (in mass range of $M_*<10^{10}\,M_{\odot}$) exhibit
metallicity distribution functions that are binomial in statistical
form i.e., the variance $\sigma(Z)^{2}$, and mean $<Z>$ are tightly
correlated, but the ratio is less than unity. Using a binomial chemical
evolution model from \citet{Oey2000}, \citet{Leaman2012} demonstrated that
galaxies approximately evolve along the $\sigma(Z)^{2} - Z$ scaling relation.  This
provides a mass-independent, self-similar framework to link two
quantities of interest: the spread in metals and the average
metallicity of a galaxy or region of a galaxy.

To further link age to these two quantities we consider the
binomial chemical evolution model of \citet{Oey2000}, which produces
metallicity distribution functions with variance and mean:
\begin{multline}
\label{eqn:nQ}
\sigma(Z)^{2} = nQ(1-Q),\\
              <Z> = nQ,\\
\end{multline}
where $n$ represents the final number of star forming
generations, and $Q$ represents the covering fraction of enrichment
events within a generation. 
To make time explicit in the model, we consider that the gas reduction increment in the \citet{Oey2000} model, $D = 1 - n\delta$, can be related through the gas fraction definition as:
\begin{equation}
\frac{M_{gas}}{M_{*}} = \frac{1 - n\delta}{n\delta}.
\end{equation} 
From this we can express an approximate star formation law and relate it to $n$ in the binomial model as:
\begin{equation}
n = \frac{M_*}{\delta} = \frac{\int_{t}^{t_{H}} \epsilon SFR(\tau) d\tau }{SFR(t)},
\end{equation}
where $t_{H}$ is the Hubble time and $t$ is when the last generation of SF happens. Following empirical and theoretical SF laws, we have introduced $\epsilon$ to allow for non-perfect conversion of gas to stars. This variable is often expressed as an inverse of the gas
depletion time: $\epsilon = 1/t_{\rm dep}$.

For a constant SFR, $n$ then becomes:
\begin{equation}
 n = \frac{t_{H} - t}{t_{\rm dep}}, 
\end{equation}
where $t_H - t$ is the length of time that all generations of star formation last in the galaxy. 
Combining this with the expressions for variance and mean Z in
equation~\ref{eqn:nQ} we find:
\begin{equation}
\sigma(Z)^{2} = Z \left(1 - \frac{Z t_{dep}}{t_{H} - t}\right). \\
\end{equation}
This can then be re-expressed as a link between age, average metallicity and spread in metallicity:
\begin{equation}
\label{eqn:azdz}
t = t_{H} - t_{\rm dep}\frac{Z}{1 - \sigma(Z)^{2}/Z}.
\end{equation}

\begin{figure}
  \centering\includegraphics[width=6.5cm]{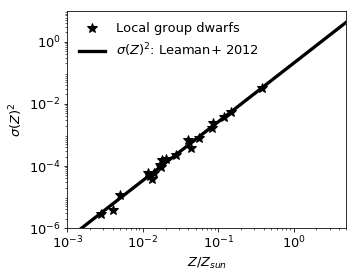}
  \centering\includegraphics[width=6.5cm]{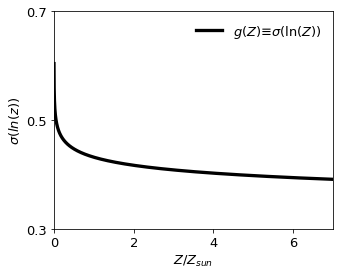}
\caption{{\bf Top: }, the relation of metallicity spread $\sigma(Z)$ vs. $Z$, the black stars are local group dwarf galaxies from \citep{Leaman2012}, the black
  curve (labeled as a) is fitting to these data points (equation~\ref{eqn:zdz}).
  {\bf Bottom: } $\sigma(ln(z))$ vs. $Z$ converted from the curve in the top based on equation~\ref{eqn:slnz}.}
\label{fig:ZsZ_Ryan}
\end{figure}

We can now use the equation~\ref{eqn:azdz} to set a mass-independent link between
age and metallicity distributions.
To further link these quantities and specify the metallicity spread in terms of average metallicity, we consider the observed statistical correlations present in metallicity distributions of Local Group dwarf to MW mass galaxies.
Empirically, the observed relation between $\sigma(Z)$ and
$Z$ from local group galaxies \citep{Leaman2012}:
\begin{equation}
  \label{eqn:zdz}
  \sigma(Z)^2 = 10^{a + b\log10(Z)},
\end{equation}
where $a=-0.689$ and $b=1.889$ shown as the black solid line in the top
panel of Figure~\ref{fig:ZsZ_Ryan}.
As our priors are best expressed in natural log space, and considering $\ln{Z}$ of each population follows a Gaussian
distribution, then a purely mathematical calculation yields
\begin{equation}
\label{eqn:slnz}
g(Z) \equiv \sigma(ln(Z)) = \sqrt{\ln(1 + \sigma(Z)^2/Z^2)}.
\end{equation}
Combined with equation~\ref{eqn:zdz}, this yields for $g(Z)$ the black solid curve shown in the bottom panel of Figure Figure~\ref{fig:ZsZ_Ryan}.

\begin{figure}
\centering\includegraphics[width=7.5cm]{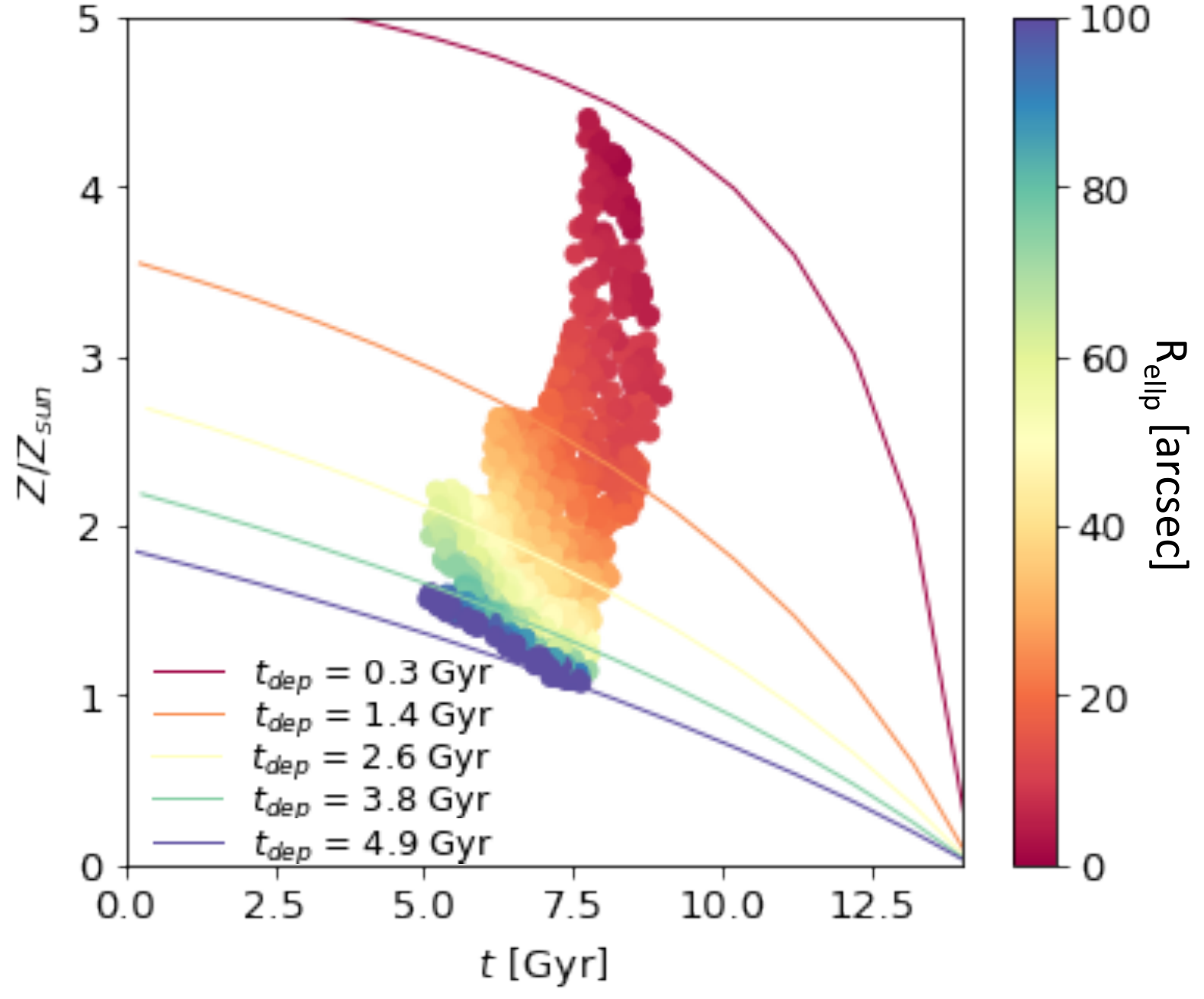}
\caption{The age-metallicity relation $Z(t | t_{\rm dep})$ derived with combination of equation~\ref{eqn:azdz} and equation~\ref{eqn:zdz}. The solid curves with different colors are $Z(t)$ by choosing different $t_{\rm dep}$ as labeled. The dots are the observed age $t_{\rm obs}$ and metallicity $Z_{\rm obs}$ (Au-6 $\vartheta=80^o$) colored by their elliptical radius $R_{\rm ellp}$ on the observational plane. }
\label{fig:agemet_Ryan}
\end{figure}

Setting $t_H = 14$ Gyr and by substituting  $\sigma(Z)^2$ from equation~\ref{eqn:zdz} into
equation~\ref{eqn:azdz}, we obtain a relation between average
metallicity $Z$ and formation time $t$. 
This age-metallicity relation $Z(t | t_{\rm dep})$ still depends on depletion time $t_{\rm dep}$. As shown in Figure~\ref{fig:agemet_Ryan}, $Z(t)$ is steeper with smaller $t_{\rm dep}$, and shallower with larger $t_{\rm dep}$.
Actually, $t_{\rm dep}$ will likely be different for different regions in a galaxy with complicated star formation history. The dots overplotted in Figure~\ref{fig:agemet_Ryan} are the observed age $t_{\rm obs}$ and metallicity $Z_{\rm obs}$ (Au-6 $\vartheta=80^o$) colored by their elliptical radius $R_{\rm ellp} = \sqrt{x^2 + y^2/(q_{\rm obs})^2}$ on the observational plane, where $q_{\rm obs}$ is observed flattening of the galaxy. There is almost a linear correlation between $t_{\rm dep}$ indicated by ($t_{\rm obs}$, $Z_{\rm obs}$) and radius $R_{\rm ellp}$ (also see figure~\ref{fig:tdepr} in the appendix). The star formation in a galaxy is consistent with smaller $t_{\rm dep}$ at small radii, and larger $t_{\rm dep}$ at large radii. We note that the range of depletion times is consistent with those found for a wide range of galaxy masses, regions - including at larger redshifts \citep[c.f.][]{Bigiel2011}.

\subsection{Bayesian analysis}
\label{SS:Bayes}
We use Bayesian statistical analysis (Python package pymc3) to obtain
age ($t_k$) and metallicity ($Z_k$) of the orbital bundles.

\subsubsection{Fit to age map}
\label{SSS:Fitt}
We first fit the age map following equation~\ref{eqn:agefit}.
In pymc3, we can specify a prior for each parameter as a distribution.
We adopt a bounded normal
distribution:
\begin{equation}
\label{eqn:tnorm}
f(t_k | \mu_k, \sigma_k) =
\sqrt{\frac{1}{2\pi\sigma_k^2}}\exp{-\frac{(t_k-\mu_k)^2}{2\sigma_k^2}}
\end{equation}
for $t_k$ with lower and upper boundary of 0 and 14 Gyr, we
set $\mu_k$ and $\sigma_k$ as follows:
\begin{equation}
  \label{eqn:t_prior}
\mu_k ={\rm Randn}(<t_{\rm obs}>,  2\sigma(t_{\rm obs}))  \hspace{9.em} 
\end{equation}
\begin{equation}
  \label{eqn:st_prior}
\sigma_k = 2\sigma(t_{\rm obs}),
\end{equation}
where $<t_{\rm obs}>$ and $\sigma(t_{\rm obs})$ indicate average and
standard deviation of age
from the observational age map. Note that ${\rm Randn}(a, b) $ means a random number
generated from normal distribution with center $a$ and dispersion $b$, the above priors are uniform for all the orbital bundles. 

Once the priors are specified, we start the Markov Chain Monte Carlo (MCMC) analysis by adopting a
student T distribution $t_{\nu}(x| \mu, \sigma)$ for the posterior likelihoods.
The chain is initialized with the method `ADVI'-Automatic Differentiation Variational Inference-with 200000 draws, and we
run 2000 steps.
We take the average and standard deviation of the last 500 steps as mean
and $1\sigma$ uncertainties of $t_k$.
The last 500 steps will also be used for
smoothing the overall age distribution of the galaxy obtained
by our model.

In general, we expect stellar kinematics to be systematically
correlated with stellar age, because stars on dynamical hot orbits are systematically older than stars on near-circular orbits \citep{Trayford2018}. From our experience, with the above priors for $t_k$, it is not easy to perfectly
recover the correlation between stellar age $t$ and orbits' circularity $\lambda_z$, especially for the
face-on galaxies (see Section~\ref{SS:Alz}).
The results could be improved by fitting a linear relation $t = t_0 + p\lambda_z \,  (\lambda_{z} \geq 0.0) $ to the $t(\lambda_z)$ relation of
the first model\footnote{For the spiral galaxies we test, our model has
relatively large uncertainty on stellar ages
of the small fraction CR orbits with $\lambda_{z} < 0.0$, thus we do not include them for the $t \sim
\lambda_z$ fit}.
Then for the second model iteration, we set the $\mu_k$ and
$\sigma_k$ of the Gaussian priors as:
\begin{multline}
  \label{eqn:t_prior2}
\mu_k = {\rm Randn}(t_0 + p\lambda_{z,k}, 2\sigma(t_{\rm obs}) - |p|/2) \hspace{1em}  (\lambda_{z, k} \geq 0.0) \\
 \hspace{1em} =t_0  \hspace{15.3em}  (\lambda_{z,k} < 0.0) \\
\end{multline}
\begin{equation}
  \label{eqn:st_prior2}
\sigma_k = 2\sigma(t_{\rm obs}).
\end{equation}
In this case, the standard deviation of $\mu_k$s are still $\sim 2\sigma(t_{\rm obs})$, similar to the previous prior.
We perform the Bayesian analysis again with the new priors.
This iterative process could be repeated more than once, but we found the results already converged after the first iteration. We stress this is only an iterative refinement on the choices of priors, not a prescribed link between age and circularity directly.

\subsubsection{Fit to metallicity map}
\label{SSS:FitZ}
After we have obtained ages of the orbital bundles, We then fit the
metallicity map following equation~\ref{eqn:Zfit}. 
Metallicity expressed in linear unit $Z/Z_{\rm sun}$ is adopted in our
analysis.
We use a bounded lognormal distribution as prior of metallicity
$Z_k$ of each orbital bundle:
\begin{equation}
\label{eqn:zlognorm}
f(Z_k | \mu_k, \sigma_k) = \frac{1}{Z_k}
\sqrt{\frac{1}{2\pi\sigma_k^2}}\exp{-\frac{(\ln{Z_k}-\mu_k)^2}{2\sigma_k^2}},
\end{equation}
with
lower and upper boundary of 0 and 10.
We first start with $\mu_k$ and $\sigma_k$ of
the lognormal distribution as follows:
\begin{equation}
  \label{eqn:z_prior}
\mu_k = \ln({\rm Randn}(<Z_{\rm obs}>,\sigma(Z_{\rm obs})))
\end{equation}

\begin{equation}
  \label{eqn:sz_prior}
\sigma_k = \sigma(Z_{\rm obs}),
\end{equation}
where $<Z_{\rm obs}>$ and $\sigma(Z_{\rm obs})$ are the average value and
standard deviation of metallicity from the observational
metallicity map.
Then we perform Bayesian analysis similar to the fitting of age map.
We take the average and standard deviation of the last 500 steps of
the MCMC chain as mean
and $1\sigma$ error of $\ln(Z_k)$, the last 500 steps are also used for
smoothing the overall metallicity distribution of the galaxy obtained
by our model.

The above uniform priors for $Z_k$ lead to a poor recovery of the
age-metallicity distribution.
To this end, we use the age-metallicity relation derived in
Section~\ref{SS:age-met} to give more reasonable priors for $Z_k$, with age of each orbit $t_k$ already obtained. We adopt again the
bounded lognormal distribution, but now with $\mu_k$ and $\sigma_k$ given by:
\begin{equation}
  \mu_k = \ln(Z(t_k | t_{\rm dep}(r_k)))  
\end{equation}
\begin{equation}
\sigma_k  = g(\mu_k).
\end{equation}
We let the depletion time locally vary as a function of radius $r_k$ (which traces mass density), and refer the reader to Appendix A for details.

In order to understand how the different priors on $t_k$ and $Z_k$ affect our results,
We perform two model fits to age and metallicity maps - an unconstrained version, and one with the above mentioned priors. These are summarized in Table~\ref{tab:prior}.
The model results from these different priors are marked as R1 and R2 respectively, throughout the paper.

\begin{table}
\caption{The priors for the Bayesian fitting to age and metallicity maps. We have $\sim 100$ orbital bundles in the model, $k$ indicates any of these. We take a bounded normal distribution as
 prior for $t_k$ (equation~\ref{eqn:tnorm}), a bounded lognormal distribution for $Z_k$ (equation~\ref{eqn:zlognorm}), with the
 mean $\mu_k$ and dispersion $\sigma_k$ specified differently for the
 two rounds of model fitting: R1, R2. When fitting to age, for Model R1, we use uniform priors for $t_k$, and for Model R2 we use a relation $t_k=t_0+p\lambda_{z,k}$ fitted from the result of model R1. When fitting to metallicity, for Model R1 we use uniform priors for $Z_k$, while for Model R2 we use the age-metallicity-metallicity spread relation $Z(t| t_{\rm dep})$ (Figure~\ref{fig:agemet_Ryan}), $g(Z)$ (Figure~\ref{fig:ZsZ_Ryan}) (see Section~\ref{SS:age-met}).}
\scriptsize\centering
\label{tab:prior}
\begin{tabular}{*{2}{l}}
\hline
  Model & prior for age $t_k$   \\
     & ${\rm Norm}(t_k|\mu_k, \sigma_k )$ \\
\hline
  R1      & $\mu_k = {\rm Randn}(<t_{\rm obs}>, 2\sigma(t_{\rm obs}))$\\
            & $\sigma_k = 2\sigma(t_{\rm obs})$     \\
  \hline
  R2     & $\mu_k = {\rm Randn}(t_0+p\lambda_{z,k}, 2\sigma(t_{\rm obs}-|p|/2))$\\
         & $\sigma_k = 2\sigma(t_{\rm obs})$ \\

\hline
\hline

 Model & prior for metallicity $Z_k$ \\
       & ${\rm LogNorm}(Z_k|\mu_k, \sigma_k )$ \\
\hline
  R1   &$\mu_k=\ln({\rm Randn}(<Z_{\rm obs}>,\sigma(Z_{\rm obs})))$\\
       &$\sigma_k =\sigma(Z_{\rm obs})$ \\
\hline       
  R2   & $\mu_k=\ln(Z(t_k | t_{\rm dep}(r_k)))$ \\
       & $\sigma_k=g(\mu_k)$\\
 \hline
 \end{tabular}
\end{table}

\section{Results on stellar orbit and population distributions}
\label{S:result}
In this section, we describe how the models match the intrinsic orbit
distribution, age-metallicity distribution and age-circularity
correlation with the nine MUSE-like mock data created from Auriga
simulation. For illustration of model fitting and some results, we do not show all
nine galaxies but just Au-6 $\vartheta = 80^o$. We refer the reader to Appendix B for results for the other galaxies.

\subsection{Best-fitting model}
\label{SS:best-fit}
\begin{figure*}
\centering\includegraphics[width=16cm]{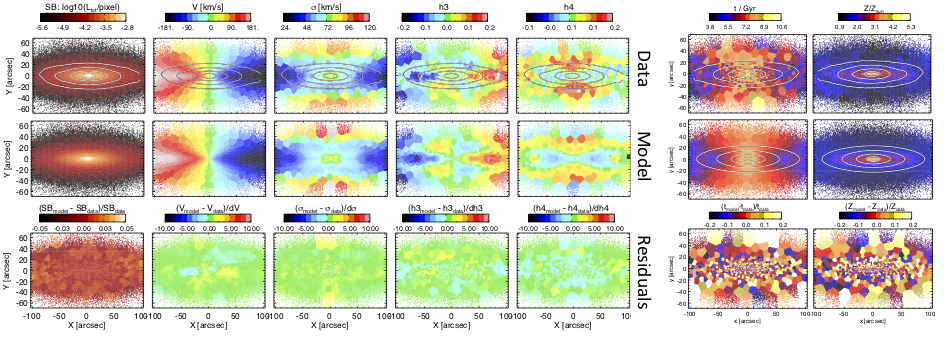}
\caption{The best-fitting model. The columns are surface
  mass density, mean velocity $V$, velocity dispersion $\sigma$,
 GH coefficient $h_3$, $h_4$, age and metallicity
  $Z$. The first row are observational data, the second row are from
  the best-fitting model, and the third row are residuals. }
\label{fig:bestfit}
\end{figure*}

A best-fitting model of the mock data from Au-6 with
$\vartheta = 80^o$ is shown in Figure~\ref{fig:bestfit}.
From left to right, the columns are surface mass density, mean
velocity, velocity dispersion, $h_3$, $h_4$, age ($t$) and metallicity
($Z$).  The first row
is the data, the second row is reproduced by the best-fitting model and the third
is residual. 
The model matches the kinematic maps, age and metallicity maps well.
For just the projected on-sky maps, we see that the models with different priors (R1, R2) fit the age and metallicity maps equally well.

In summary, up to this point we have obtained an orbit-superposition model, with the
orbit weights solved by matching the stellar mass distribution and
kinematic maps. Here we further divided the orbits into $\sim 100$ bundles,
and obtained the age and metallicity of these bundles
by fitting the age and metallicity maps. By taking a Bayesian
statistical analysis, we obtained the mean value $t_k$ and $Z_k$ of each bundle $k$, as well as their uncertainties $\sigma(t_k)$, $\sigma(Z_k)$.

\subsection{Stellar orbit distribution}
\label{SS:lz_dis}

\begin{figure*}
\centering\includegraphics[width=14cm]{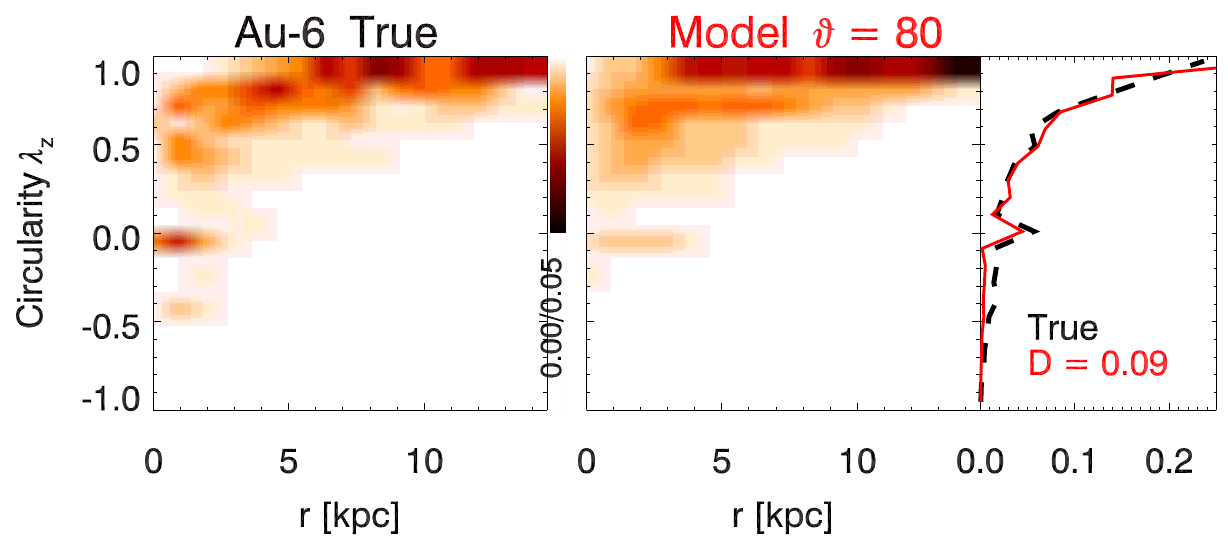}
\caption{The stellar orbit distribution described as probability
  density of orbits in the phase space of
  $\lambda_z$ vs. $r$, for Au-6.
  The left panel is the true stellar orbit
  distributions from simulations. The second
   is the distribution of orbits in our
best-fitting model for mock data with $\vartheta = 80^o$. The right panel is comparison of marginalized $\lambda_z$ distribution between true and model. Similar figures for the other galaxies are included in the appendix Figure~\ref{fig:rlz_all}.}
\label{fig:halo6_rlz}
\end{figure*}

We first check how well the orbit distribution in our model matches
the true distribution from the simulation. 
The real gravitational potential and 6D phase-space information of particles are known
in the simulation. Thus we know the instantaneous
circularity $\lambda_z$ of each particle
\citep{Facundo2017}, which does not necessarily conserve $\lambda_z$
when orbiting in the potential, especially for those particles on
radial/box orbits with $\lambda_z \sim 0$.
To obtain the orbits' circularity, in principle, we have to freeze the
potential, integrate the particle orbits in the potential, and calculate the
average values along the orbits. Here for simplicity,
we use a single snapshot and select those particles that are close in energy, $E$,
angular momentum, $L_z$ and the total angular momentum amplitude, $|L|$. Under the
assumption that these particles are on the same orbit in a near axisymmetric system,
we then compute the corresponding averages of radius $r$ and circularity $\lambda_z$ of
these particles, which are taken as the orbit's $r$ and $\lambda_z$.
The stellar orbit distribution of one
galaxy is then presented as the probability
density distribution of all these orbits in the phase space of  $r$
vs. $\lambda_z$, which is shown in 
the left panel of Figure~\ref{fig:halo6_rlz} for Au-6. 

In our model, we calculate orbit's circularity and time-averaged radius
from the particles sampled from the orbit with equal time interval.
The middle panel of Figure~\ref{fig:halo6_rlz} shows the the distribution of orbits in our
best-fitting model for mock data Au-6 $\vartheta = 80^o$.
 
Our model matches the major features in the phase-space of $r$
vs. $\lambda_z$ as the true orbit distribution from the simulations. For
the case of Au-6 $\vartheta=80^o$ we show here, counter-rotating (CR) orbits contribute a small fraction in the simulation, and our model underestimate CR orbits by $\sim 50\%$.
The right sub-panel is the marginalized $\lambda_z$ distribution. The black
dashed curves is the true distributions; red solid curves represent
that from our model.
We did a 1D KS test to check how well the $\lambda_z$ distribution
recovered by our model match the true distribution from simulation. The
D-statistics $D$ is the
maximum deviation from the accumulated curves of two distributions.
We obtained $D = 0.09$ here for the $\lambda_z$ distribution
A similar comparison for Au-5, Au-6, Au-23 with inclination angles of
$\vartheta = 40^o, 60^o, 80^o$ are shown in the appendix in Figure~\ref{fig:rlz_all}.

\subsection{Age-metallicity distribution}
\label{SS:AM_dis}
\begin{figure*}
\centering\includegraphics[width=17cm]{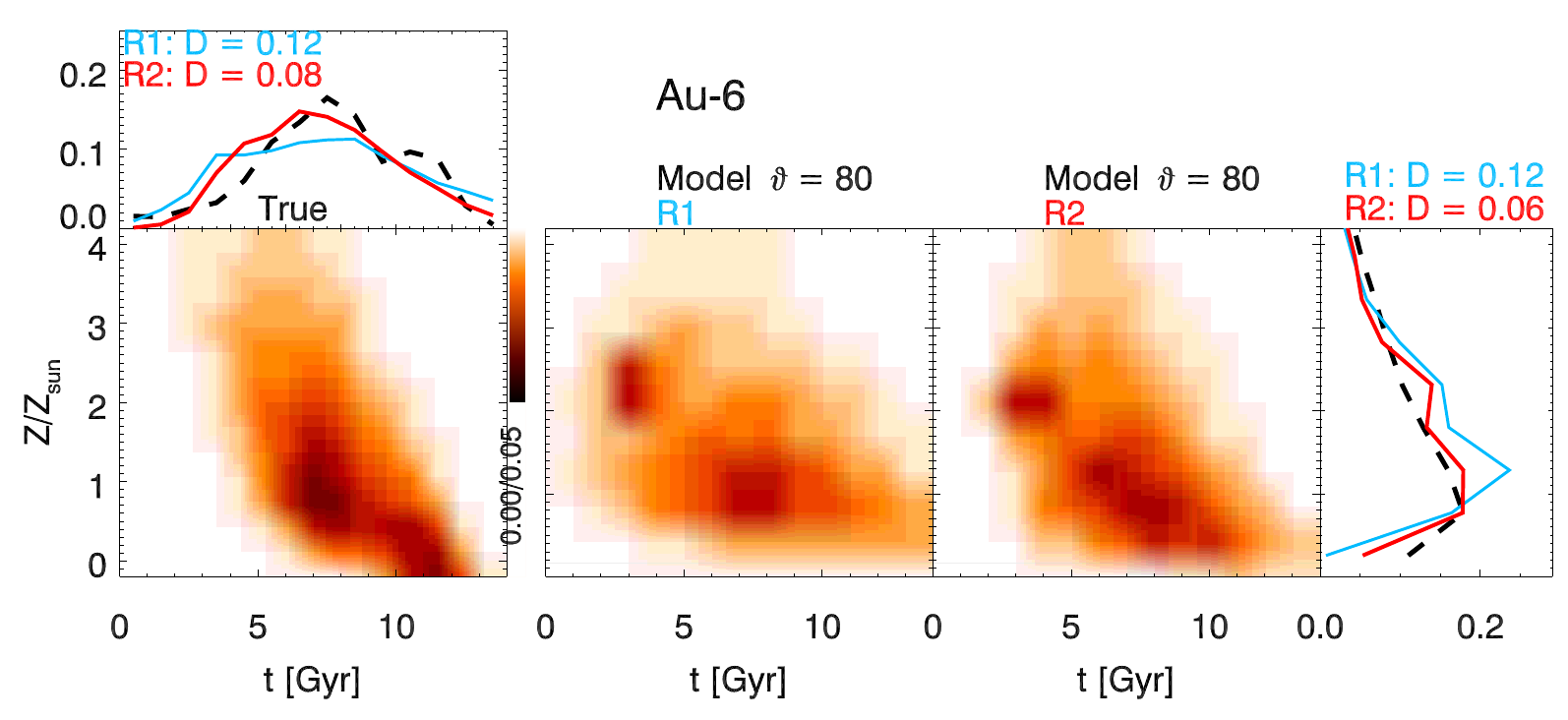}
\caption{The intrinsic age-metallicity distribution for Au-6
  $\vartheta = 80^o$. 
  The left panel is the true distribution in stellar age $t$
  vs. metallicity $Z$ of particles of the
  simulation. The rest panels are obtained by our model constrained by
  mock data Au-6 $\vartheta = 80^o$ with priors R1 and R2 from left
  to right.
  The probability contours of our models are smoothed by the last 500
  steps of MCMC chains of $t_k$ and $Z_k$ from the Bayesian analysis. 
  The upper subpanel is the marginalized age distribution and the
  right sub-panel is the marginalized metallicity distribution. The black
dashed curve is the true, red, blue solid curves represent
that from model with priors R1, R2, respectively. The D-statistics $D$
calculated from KS test comparing 1D age/metallicity
distribution from our model to the true age/metallicity distribution are labeled.}
\label{fig:halo6_tZ}
\end{figure*}

Age and metallicity maps projected on-sky can be reproduced with many
degenerate combinations of age-metallicity distributions of the stars.
However, not all combinations may be physical, nor match the intrinsic
age vs. metallicity distribution of the simulated galaxy.
Here we check how the age and metallicity distribution of orbits
in our models match the intrinsic distribution of particles in these simulations.

In Figure~\ref{fig:halo6_tZ}, we show the probability density distribution
of particles/orbits in age ($t$) vs.
metallicity ($Z/Z_{\odot}$), from the simulations and from our model of Au-6 $\vartheta = 80^o$.
The first panel labeled with `True' shows the true distribution on Age vs. $Z$ of particles in
 the simulation. The following panels are those obtained by our model for
 mock data $\vartheta = 80^o$ but with prior R1, R2 from left to right.
 The probability contours are smoothed by the last 500 steps of MCMC
 chains of $t_k$ and $Z_k$ from the Bayesian analysis.

 The upper sub-panel for each halo is the marginalized age distribution and the
 right sub-panel is the marginalized metallicity distribution. The black
dashed curves are the true distributions; red, blue solid curves represent
those from models with prior R1 and R2, respectively.
From a 1D KS test, we obtained $D = 0.12, 0.08$ for age distribution and $D = 0.12,
0.06$ for metallicity distribution, for models with prior R1 and
R2, respectively. Both intrinsic age and metallicity distributions are recovered better with model R2 than R1.

In the true distribution, most
stars follow a relation with older stars that are more metal-poor.
Model R1 hardly recovers this
relation (Figure~\ref{fig:halo6_tZ}), missing a significant fraction in mass of sub-solar metallicity stars, and showing roughly uncorrelated distributions of constant metallicity groupings over a wide range in age.
The recovery of age-metallicity relation significantly improved with model R2, especially for more face-on galaxies (see
Figure~\ref{fig:tZ_all_noprior} and Figure~\ref{fig:tZ_all}).

\subsection{Age-circularity correlation}
\label{SS:Alz}
In this Section, we study the correlation of stellar orbit circularity and ages in the simulation, and
check how well the correlation can be recovered by our models.

\begin{figure*}
\centering\includegraphics[width=17cm]{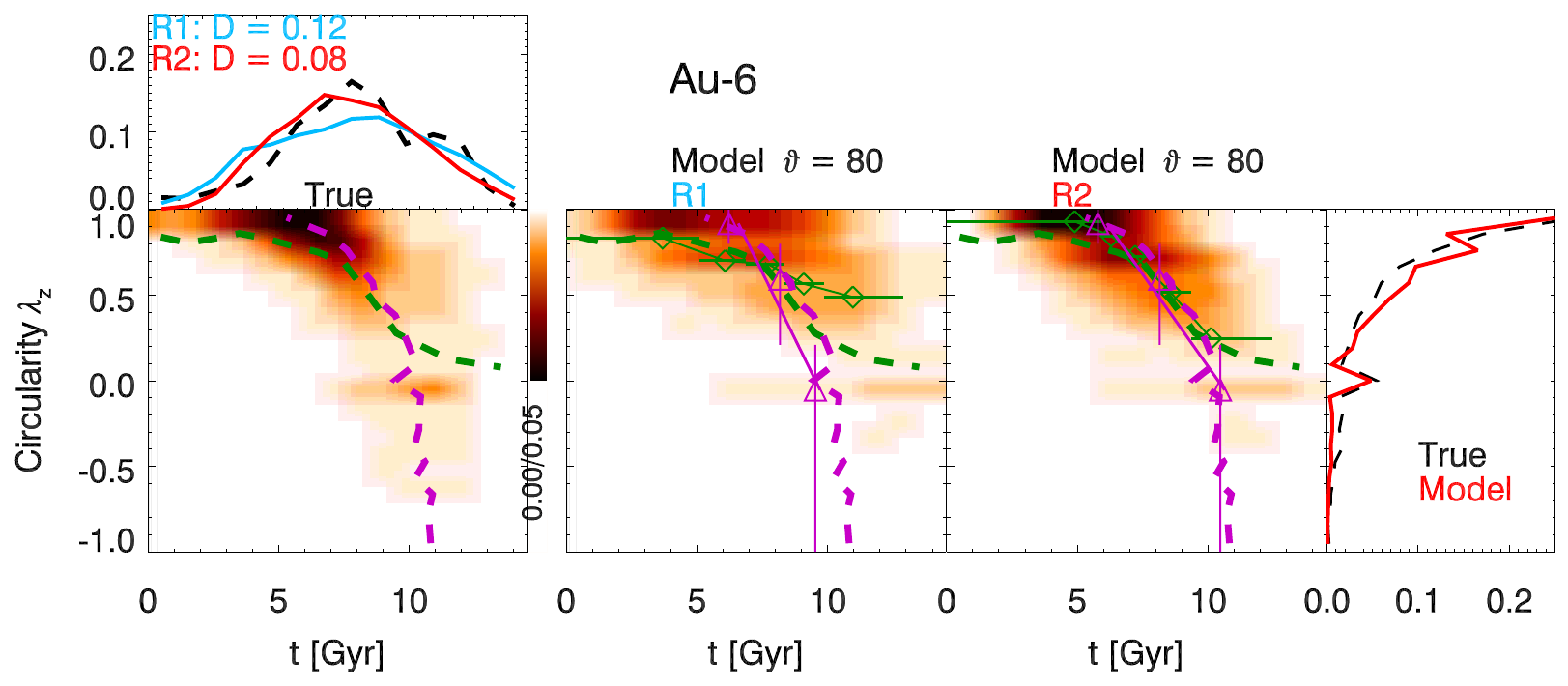}
\caption{The intrinsic correlation of age vs circularity, for Au-6 and
  those from our models with mock data $\vartheta = 80^o$.  The first
  one is the true
  distributions of the particles from simulation, darker color
  indicates higher probability density, the magenta dashed
curve is average age $t$ as a function of $\lambda_z$, while the green dashed curve is average $\lambda_z$ as a function of age $t$ from the
simulation. The following panels are obtained by model R1 and R2
respectively. In each panel, the
magenta triangles are average age as a function of $\lambda_z$ for the corresponding panel,
the magenta solid line is a linear fit ($t = t_0 + p \lambda_z$) to the triangles. The magenta solid line from
model R1 is used as priors of ages $t_k$ in model R2. Similarly, the green diamonds represent average $\lambda_z$ as a function of age $t$ for the model. Model R2 matches the true relations in the simulation better than model R1. }
\label{fig:halo6_tlz}
\end{figure*}

The intrinsic probability density distribution of orbits
on age $t$ vs. circularity $\lambda_z$ for simulation Au-6 is shown in the left panel of Figure~\ref{fig:halo6_tlz}. 
darker color indicates higher probability
density. In the simulation, there is a correlation between stellar age and orbits' circularity: highly circular orbits are systematically younger,
and radial-motion dominated orbits are older. 
We calculate the average
age of orbits as a function of $\lambda_z$ by binning on $\lambda_z$ (the magenta dashed curve) and average $\lambda_z$ as a function of age by binning on age $t$ (the green dashed curve).

The orbit distribution on age vs. circularity obtained by
our models with Au-6 $\vartheta = 80^o$ are shown in the
following panels, for model R1 and R2 respectively.
In each panel, the probability contours represent the distribution that
is derived from the last 500 steps of MCMC chain of $t_k$ from the Bayesian analysis. The magenta triangles are average age as function of $\lambda_z$ from the model.
The magenta solid line ($t = t_0 + p\lambda_z$) is a linear fit to the triangles, which from model R1 is used as prior of age $t_k$ for model R2 when fitting to age map. The green diamonds are average $\lambda_z$ as a function of age by binning on age from the model. 

Our models generally match the $t \sim \lambda_z$ correlation from the
simulation, model R2 matches it better than R1 for Au-6
$\vartheta=80^o$; the improvement of R2 comparing to R1 is more significant for
more face-on galaxies (see Figure~\ref{fig:tlz_all_noprior}, Figure~\ref{fig:tlz_all}). 
The results in following sections are based on Model R2 if not otherwise specified.

\section{Orbital decomposition}
\label{S:decomp}
To further quantify the correlation between the orbits' dynamical
properties and stellar populations, we decompose galaxies \citep[c.f.,][]{Zhu2018a} by dividing the orbits into cold ($\lambda_z \geq 0.8$), warm ($0.25 \leq \lambda_z<0.8$),
hot ($|\lambda_z|<0.25$), and counter-rotating (CR, $\lambda_z<-0.25$)
components.
We emphasis that the separation of cold, warm, hot+CR components is just for proof of concept.  For real galaxies, we may adjust the component separation case by case.

\begin{figure*}
\centering\includegraphics[width=17cm]{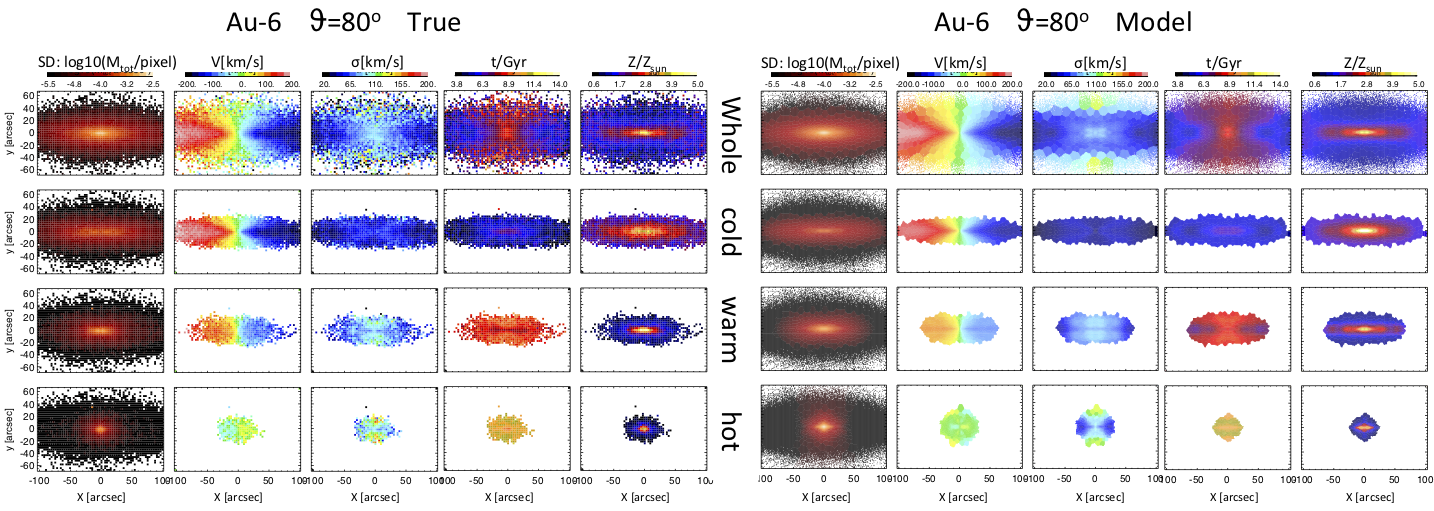}
\caption{Surface mass density, age, and metallicity maps of the
  whole galaxy,  cold, warm, and hot + CR component (from top to bottom) of Au-6 $\vartheta =
  80^o$. The left panels are constructed by particles in the simulation, the right panels are constructed
  by orbital bundles in our model constrained by mock data with $\vartheta =
  80^o$. The galaxy is at a distance of 30 Mpc, $1'' = 145 \, {\rm pc}$.}
\label{fig:halo6_ztmaps_kin3}
\end{figure*}

\begin{figure}
\centering\includegraphics[width=7cm]{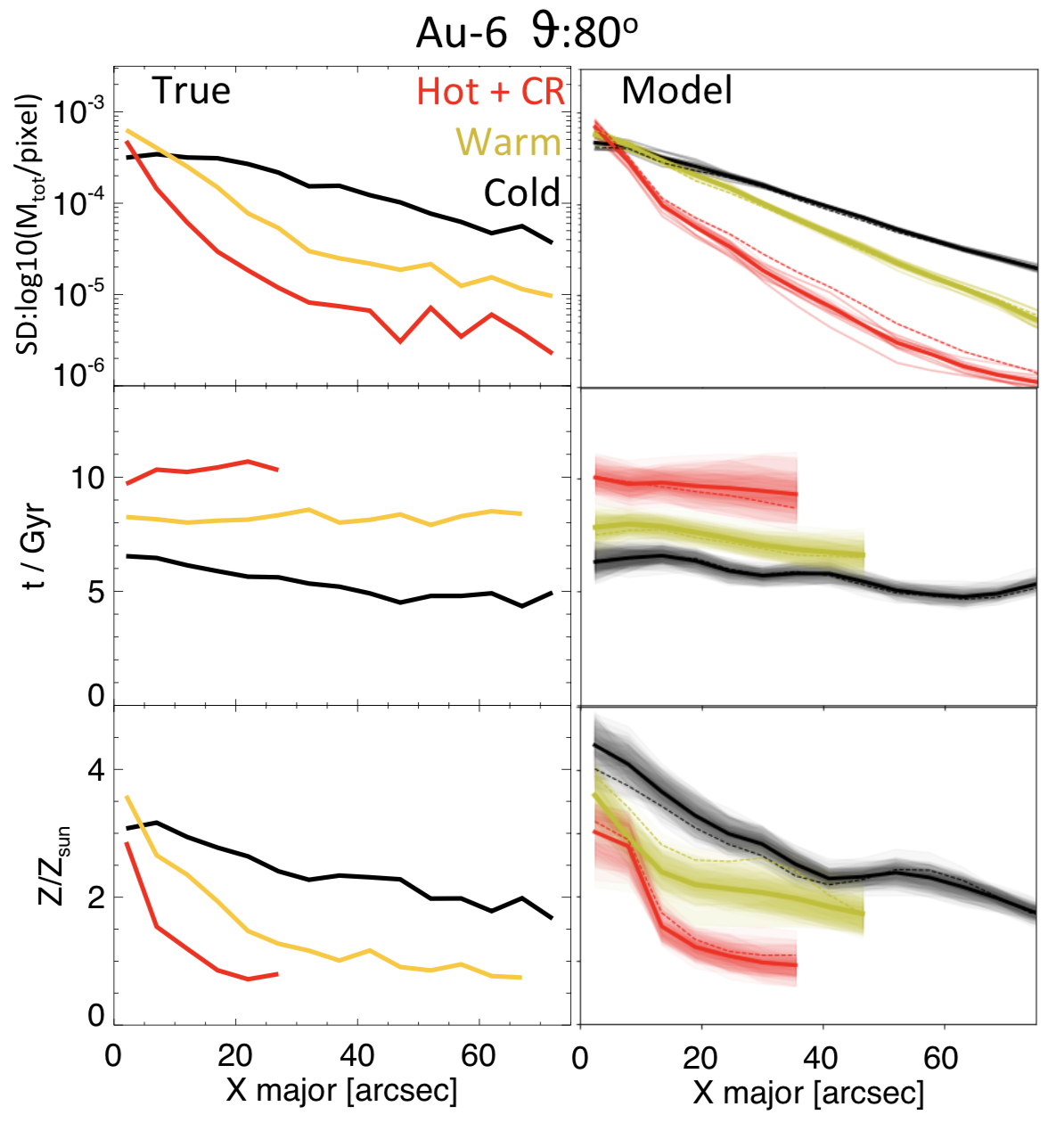}
\caption{Surface mass density,
  age, and metallicity profiles along major axis ($1'' = 145 \, {\rm pc}$) of the cold (black), warm (warm), and hot + CR (red) component of Au-6 $\vartheta =
  80^o$. The left panels are constructed by particles in the
  simulation. The right panels are those constructed
  by orbital bundles in our model constrained by mock data with $\vartheta =
  80^o$, the shadow regions indicate $1\,\sigma$ uncertainty of our models, the solid thick curves are average of models within $1\,\sigma$, and the dashed thin curves are the best-fitting one. }
\label{fig:halo6_zt_kin3}
\end{figure}

We rebuild the 3D structure for each of the cold, warm, hot + CR
components, by particles in simulations and orbits in models. Then we project the 3D
structures, here with the same inclination angle as the galaxy was observed, to
the observational plane, thus obtaining surface density (SD), mean velocity and velocity dispersion, age, and metallicity maps
for each component.
In Figure~\ref{fig:halo6_ztmaps_kin3}, we compare
these maps from the simulation Au-6 (left) to those recovered from
our model with mock data Au-6 $\vartheta = 80^o$ (right).

 Our model generally reproduces the morphology, kinematics, age and metallicity maps of the different components: the cold component is a thin disk, spatially extended, fast rotating with small velocity dispersion, young and metal-rich; the warm component is thicker and less radially extended, with weaker rotation and higher velocity dispersion, older and metal-poorer; and the hot + CR component is spheroidal and spatially concentrated, with almost no rotation and high velocity dispersion, with oldest and most metal-poor stellar populations. The 2D maps, both along the major and minor axis, of each component are visually well recovered by our model.

For a quantitatively comparison, we
show in Figure~\ref{fig:halo6_zt_kin3} the radial profiles (along the major axis) of the SD, age and
metallicity for the cold, warm, hot + CR
component, obtained from the simulation (left) and from our model (right).
The three components are plotted as black, yellow and red. In the right panels, we show not only the best-fitting model, but all the models within $1\,\sigma$ uncertainty when fitting to kinematics \citep{Zhu2018a}. The shadow areas indicate the scatter of these models within $1\,\sigma$ uncertainty, the solid thick curves are corresponding averages and the thin dashed curves are the best-fitting one.

The simulation shows an increase in stellar age from cold to
hot orbits, with little change in that behaviour with galactocentric
distance; there is only a shallow negative gradient for the cold disk
from inner to out regions. 
Our models generally reproduce this behaviour.
An implication of this is that for the galaxy as a whole, the projected
age gradient is a result of different dynamical components
super-imposed: the old-hot component dominates in the centre and a
young-cold component dominates in the outer regions.

The three components have similar metallicity ($Z/Z_{\rm sun}$) at the
center, with a strong negative metallicity gradient in the hot+CR component, and the gradient becomes weaker from hot+CR, warm to cold component. 
Our models generally match the metallicity gradients for warm and hot+CR component, but over-estimate
the metallicity of cold component in the inner region, thus resulting in a too strong metallicity gradient for the cold component.

A similar decomposition is performed for all galaxies.
For edge-on galaxies ($\vartheta=80^o$),
the age and metallicity profiles of three components are
recovered similarly well for Au-5, Au-6 and Au-23 (see
figure~\ref{fig:halo523_ztmaps_kin3}, figure~\ref{fig:halo523_zt_kin3}).
Age and metallicity profiles of each component are recovered less well in more face-on
galaxies.

\section{Global Age-dispersion relation}
\label{S:tsigmaz}
The stellar age vs. vertical velocity dispersion $\sigma_z$ relation is widely used for resolved systems to study the dynamical heating processes \citep[e.g.,][]{Leaman2017}. Here, we extract similar relations for external galaxies based on our model to galaxies with integrated-light data. Application of a similar approach to NGC 3115 has provided a $t \sim \sigma_z$ relation of this galaxy \citep{Poci2019}. Here we check how reliable the global (not disk alone) $t \sim \sigma_z$ relation can be recovered.

We can construct $t \sim \sigma_z$ relations by separating the galaxies into multiple components in two ways based on Figure~\ref{fig:halo6_tlz}, by applying a cut either on circularity $\lambda_z$, or on stellar age of the orbits in our model. 

First we follow the separating on circularity $\lambda_z$ as we did in last section, to separate the simulation/model into
cold, warm, hot+CR components, then we calculate the average age and $\sigma_z$ of each component.
In Figure~\ref{fig:tsigmaz_5t}, we show the resulting $t \sim \sigma_z$
relation in these simulations and how our model recovered it. 
The three panels are for Au-5, Au-6, Au-23,
respectively. In each panel, the black asterisks are the true ages and
velocity dispersions $\sigma_z$ of each component from the
simulation. 
There are strong age vs. $\sigma_z$ correlation in these
three Auriga simulations; cold components have small $\sigma_z$ and
are younger, while hot components have larger $\sigma_z$ and are older.

The red, blue, and purple diamonds represent those calculated from our models
for galaxies with $\vartheta = 80^o,60^o,40^o$, respectively.  
For all three simulations, our models match the average age of each orbital component well, thus also the $t \sim \sigma_z$ correlation.
There are slightly larger offsets for the face-on galaxies ($\vartheta = 40^o$), but they still generally
match the trend. Our method works better for Au-5 and Au-23, in which
the intrinsic age-$\sigma_z$ correlations are steeper, than Au-6, in which the correlation is shallower.

\begin{figure}
\centering\includegraphics[width=8.2cm]{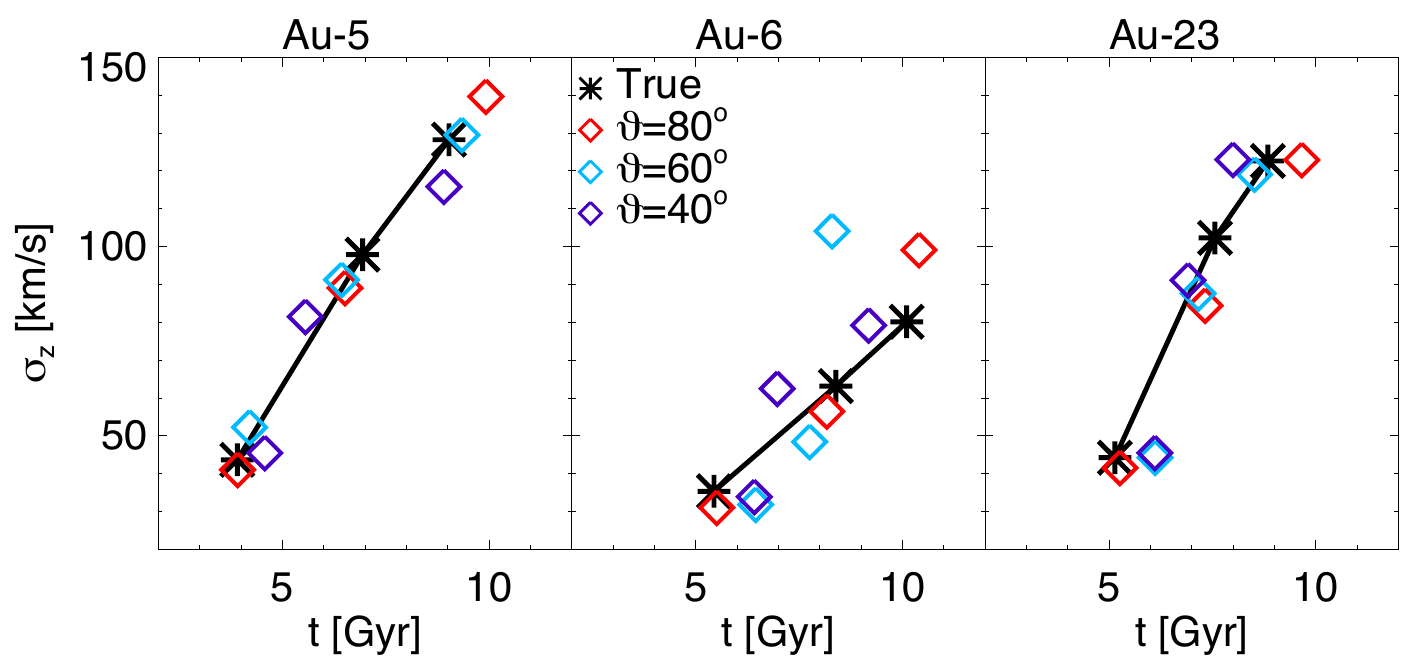}
\caption{The global age $t$ vs. dispersion $\sigma_z$ relation. The three panels are for Au-5, Au-6, Au-23, respectively. 
we separate each galaxy to be cold, warm, hot+CR components based on the orbits' circularity $\lambda_z$, and calculate average age and dispersion $\sigma_z$ for each component.
In each panel, the black asterisks are the true values calculated from the simulations, dispersion increases with age from cold, warm to hot component, the solid black curve just connects the asterisks. 
The red,
blue, and purple diamonds represent those calculated from our models
for galaxies with $\vartheta = 80^o,60^o,40^o$, respectively. }
\label{fig:tsigmaz}
\end{figure}

As local observations have traditionally computed the velocity dispersion of stars in similar age bins, we also separate the galaxy by applying cuts on stellar age, with equal mass in each bin. We use 10 age bins for the simulation, and 5 age bins for the models, and calculate average age and dispersion in each bin. In this way, it can be compared to similar observed vertical dispersion of galaxies at high redshift. The resulting $t\sim \sigma_z$ relation is shown in Figure~\ref{fig:tsigmaz_5t}. 

By binning along stellar age, our models still recover the $t\sim\sigma_z$ relation reasonably well for edge-on galaxies. It is recovered less well for face-on galaxies, for which $\sigma_z$ of old populations are under-estimated by our model. This is likely due to the relative large uncertainty of age of each orbital bundle. 
Some cold orbits could get old ages, and so contaminate the old populations and lead to an under-estimation of $\sigma_z$. 

\begin{figure}
\centering\includegraphics[width=8.2cm]{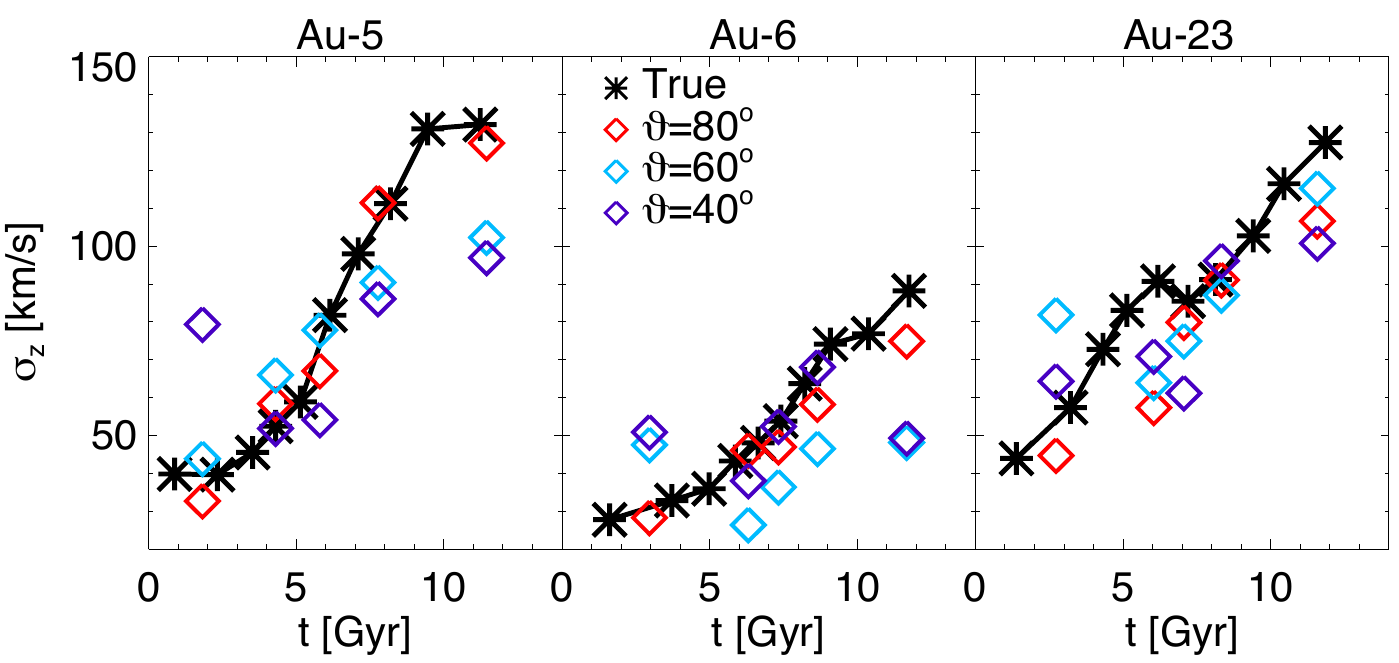}
\caption{The global age $t$ vs. dispersion $\sigma_z$ relation. Similar to Figure~\ref{fig:tsigmaz}, but by separating the simulation/model in to equal mass bins according to stellar age. We separate the simulation into 10 age bins, and model into 5 age bins.}
\label{fig:tsigmaz_5t}
\end{figure}

\section{Discussion}
\label{S:dis}
We have shown that our population-orbit superposition methods work well in recovering the intrinsic stellar orbit distributions and stellar population distributions of external galaxies.
This method could be widely applied to nearby galaxies with IFU observations, making it possible to separate structures in external galaxies from a combination of stellar kinematics and stellar chemical properties, thus bridging the gap between the Milky Way and external galaxies. 

The current method works well in a few important aspects, but also as we have shown, the interpretation of some results need to be taken with caution as it does not work equally well for all projections. Here we discuss in detail some limitations and how to improve it in the future.

\subsection{Features of bars}

\begin{figure}
\centering\includegraphics[width=7.5cm]{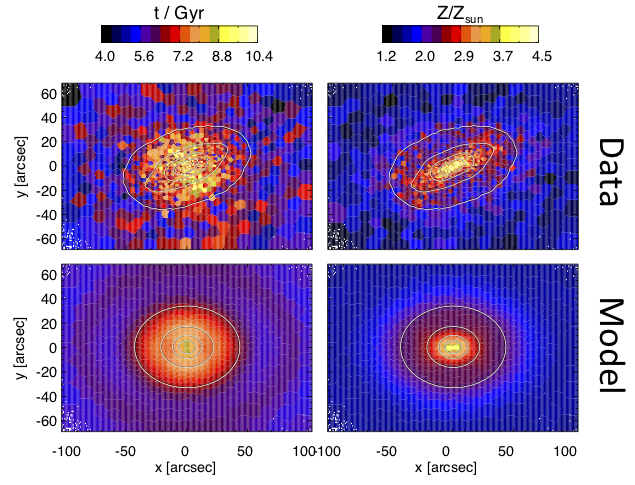}
\caption{The best-fitting model of age (left) and metallicity (right) maps of Au-23
$\vartheta = 40^o$. The first row is the mock data, a bar-like
structure is not obvious in age map, but significant
in metallicity, the contours
over-plotted illustrate the real surface
mass density of the galaxy which has a strong barred structure. The
second row is our model fitted age and metallicity maps, the
over-plotted contours illustrate the surface mass density of our
model. }
\label{fig:halo23_i40}
\end{figure}

We do no have a bar structure explicitly in the model.
While Auriga galaxies are strongly barred.
The bar regions of these galaxies are
filled by mostly warm orbits with similar circularity in our model as the
resonant orbits supporting the real bar in the simulation.
Bars generally have similar stellar
age as the disks, but are metal richer. We take Au-23 $\vartheta = 40^o$ shown
in Figure~\ref{fig:halo23_i40} as an example. The first row shows the
mock data of age and metallicity maps with contours showing the real surface mass density. The second row is our best-fit to the data with contours showing surface mass density in our model.
The bar is not a prominent feature in the age map, but much more obvious in
the metallicity map.
Based on the orbital constructions in our model, we do not have the ability to match the bar structure in the metallicity map. This could directly lead to a bias in the recovered metallicity for
different orbital components.

For edge-on cases, the structure in metallicity caused by the bar
could be roughly matched by assigning different metallicities to the
corresponding warm orbits, thus our model can still work on recovering
metallicities of cold, warm, hot+CR components. This is not the
case for face-on projections.
Including a bar explicitly in our Schwarzschild model in the future, as attempted in other studies \citep{Vasiliev2019}, will certainty
lead to improving recovery of metallicities of different structures in
barred galaxies. 
In such a model, we may need a third parameter to characterise orbits, besides radius $r$ and circularity $\lambda_z$. Then orbit bundles divided on a 3D phase-space rather than 2D $r\sim\lambda_z$ plane might be used for tagging age and metallicities.

\subsection{Beyond single age and metallicity per orbit}
We tag a single value of age and metallicity to each orbit bundle divided in the 2D $r\sim\lambda_z$ plane, while each orbit bundle should have a distribution of age and metallicities. 
A consequence of this is the most-poor end of the metallicity distribution is difficult to match completely (See Figure~\ref{fig:halo6_tZ} and Figure~\ref{fig:tZ_all}). This can be due to two effects which we explain below.

\begin{figure}
\centering\includegraphics[width=8cm]{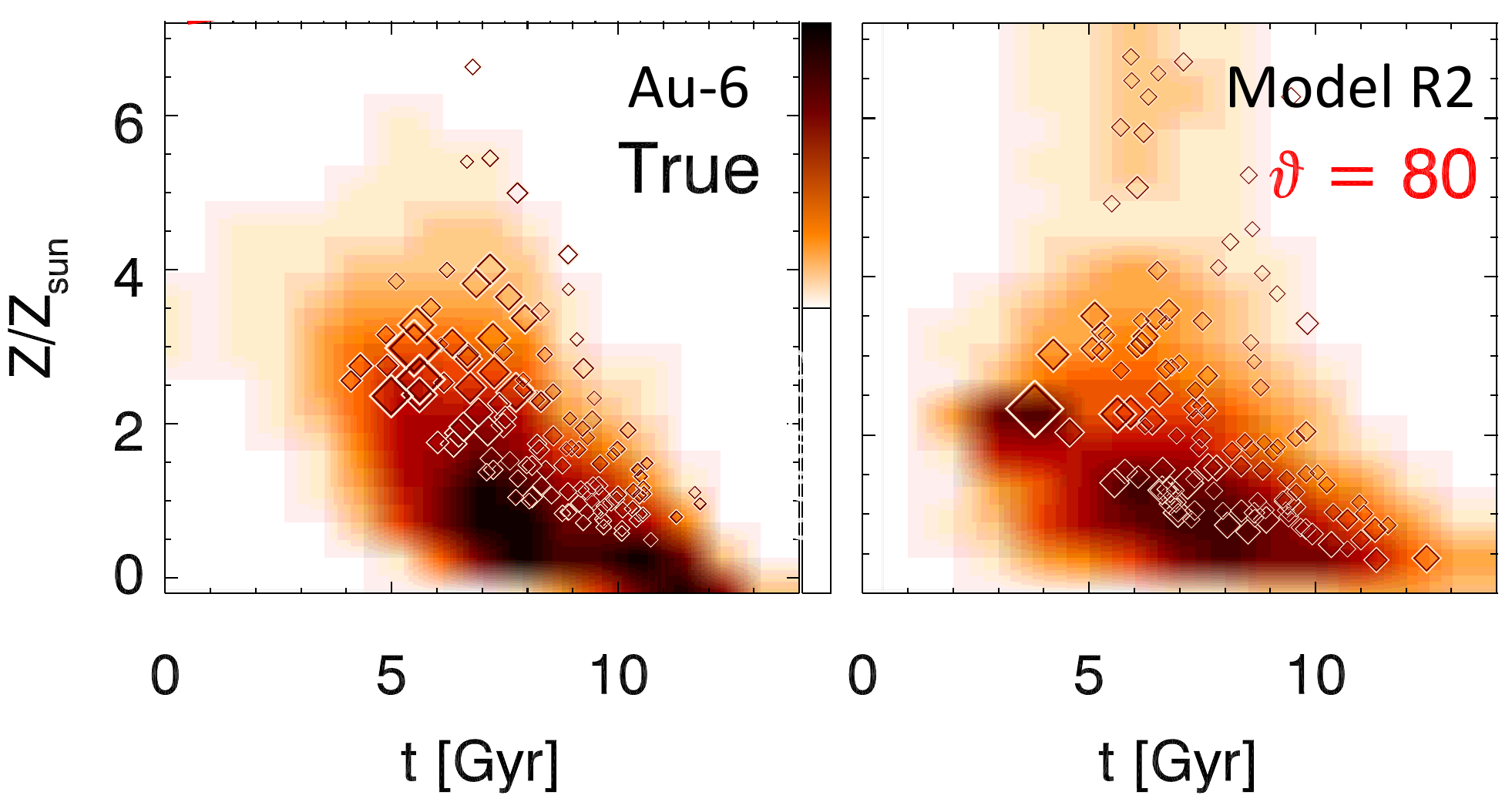}
\caption{The intrinsic age-metallicity distribution for Au-6 $\vartheta = 80^o$. In the left panel, the contours represent the probability density distribution of particles in the simulation, the diamonds represent age vs. metallocity of orbital bundles, which we obtained by dividing the particles based on 2D $r\sim\lambda_z$ plane. In the right panel, the diamonds are age vs. metallicity of orbital bundles in our model, the contours are smoothed by results of the last 500 steps of the pymc3 process.}
\label{fig:halo6_otz}
\end{figure}

In the left panel of Figure~\ref{fig:halo6_otz}, the contours are probability density distribution in age vs. metallicity of particles in the simulation. We divide the particles into different orbital bundles on $r\sim\lambda_z$, each diamond represent average age and metallicity of an orbit bundle. As can be seen, the age and metallicity distributions of the orbit bundles are narrower than those of true distribution of particles. There are rarely orbit bundles with average $Z/Z_{\rm sun} < 0.5$ - even in the simulation. Thus it is expected that when assigning a single value of age and metallicity to each orbital bundle, our model will also show a narrow distribution in age and metallicity (even smoothing over the last samples of our pymc3 process).

A second, related aspect, is that our model is reproducing on-sky projected age and metallicity maps. For any projection, even at the high spatial resolution of modern MUSE observations, such spatial binning results in a significant loss in information when compared to the true particle age and metallicities. Further work looking at optimal reconstruction of true particle distributions from binned maps and observational estimates of line-of-sight metallicity and age distributions per pixel will provide help in this front.
Technically, it is not difficult to impose an age and metallicity distribution to each orbital bundle. However, the distribution is fully unconstrained by our current data, which are only light/mass weighted age and metallicity maps averaged along line-of-sight.
If we want to constrain the age and metallicity distributions of each orbital bundle, we will need line-of-sight age and metallicity distribution from observation, which we still need to further investigate from the observational side. 

We find that the method works better for edge-on than face-on galaxies in a few
aspects: recovering the general age vs. circularity correlation, the detailed age and metallicity profiles of different dynamical components, and the $t \sim \sigma_z$ relation. Apart from the presence of bars, age and metallicity information of different structures, e.g, thin/thick disks and bulge, are revealed in the edge-on age/metallicity maps, while blended in face-on projected data.
The ability of recovering those properties for face-on galaxies could also
improve if we can use line-of-sight age/metallicity distribution from observations as model constraints.
 
\section{Summary}
\label{S:summary}
We present a population-orbit superposition method in this paper by tagging age
and metallicity to orbits in the Schwarzschild model and requiring it to fit the observed luminosity/mass
distribution, as well as stellar kinematics, age and metallicity maps. 
We validate the method by testing against mock data created from simulations.
We take three simulations from Auriga, and project each simulation with three different inclination angles $\vartheta = 80^o, 60^o, 40^o$. With each projection, we create a
set of mock data with MUSE-like data quality, including surface mass density, stellar kinematics, age and metallicity maps.
Thus, we have nine mock datasets in total, each is taken as an independent observed
galaxy, to which we apply our method. 

The mock data is fitted well by our model
with no difficulty except for the barred features in face-on galaxies. To reproduce correct relations between age and metallicity, we found a physically motivated chemical evolution prescription for the priors significantly improved the results.
To evaluate the method's ability of recovering galaxies' intrinsic
properties, we compare these properties from our models to those from simulations:
\begin{itemize}
\item[(1)] Our models can generally and equally well recover the stellar
  orbit distribution in the
  phase-space of circularity $\lambda_z$ vs. radius $r$ for galaxies
  with different viewing angles.

 \item[(2)] The intrinsic stellar population distribution in age $t$
   vs. metallicity $Z$ is hard to fully recover. We derived a
   theoretically motivated link between age, mean metallicity and metallicity
   spread, which we impose as
   priors when fitting metallicity maps.
   This link improved
   our recovery of age-metallicity correlations, and the marginalized
   metallicity distributions. 

 \item[(3)] Our method works well in recovering the age-circularity correlation for edge-on
   galaxies, but less well for more face-on galaxies. An
   iterative fitting by updating the priors for age based on an initial fit helps improving the results, especially for
   face-on galaxies. 

\item[(4)] To further check the method's ability on recovering intrinsic properties of different galaxy structures: we decompose galaxies into cold ($\lambda_z>0.8$), warm
    ($0.25<\lambda_z<0.8$), hot + CR ($\lambda_z < 0.25$) components. We
    then rebuild the surface density, mean velocity, velocity dispersion, age and metallicity maps of each component.
    By comparing with those constructed from the simulation, we find these maps of each component are quantitatively well recovered by our model for projections close to edge-on.
   
  \item[(5)]  All three simulations have a strong
    global age($t$) vs. velocity dispersion ($\sigma_z$) correlation such that older stars are hotter with larger $\sigma_z$.
    This relation is well recovered by our
    method for all galaxies with different projection angles when we bin on circularity:
    they become older and with larger $\sigma_z$ from cold, warm to hot components. When we bin on stellar age, the $t\sim\sigma_z$ relation is still recovered reasonably well for edge-on galaxies, but we under-estimate $\sigma_z$ of old populations for face-on galaxies.
 \end{itemize}

The results presented will be our basis to apply this method to real data, including case/statistical studies for galaxies with MUSE-like IFU observations. The decomposition of cold, warm, hot/CR components is not a final solution for dynamical decomposition of real galaxies, as flexible choice for galaxies case-by-case could be investigated. While continued improvements to the methodology will be developed by our team, this proof-of-concept shows great promise in the ability of the method to uncover the buildup and timescales for formation of different components within galaxies observed with modern IFU instruments. 

\bibliography{chemo1.bib}

\section*{Acknowledgments}

This research is supported by the National Key R$\&$D Program of China under grant No. 2018YFA0404501 (SM, LZ), and National Natural Science Foundation of China under grant No. Y945271001 (LZ). GvdV and PJ acknowledges funding from the European Research Council (ERC) under the European Union's Horizon 2020 research and innovation programme under grant agreement No 724857 (Consolidator Grant ArcheoDyn). J. F-B acknowledges support through the RAVET project by the grant AYA2016-77237-C3-1-P from the Spanish Ministry of Science, Innovation and Universities (MCIU) and through the IAC project TRACES which is partially supported through the state budget and the regional budget of the Consejer\'\i a de Econom\'\i a, Industria, Comercio y Conocimiento of the Canary Islands Autonomous Community. RMcD is the recipient of an Australian Research Council Future Fellowship (project number FT150100333).


\appendix
\section{Depletion time in metallicity priors}
\begin{figure*}
\centering\includegraphics[width=14cm]{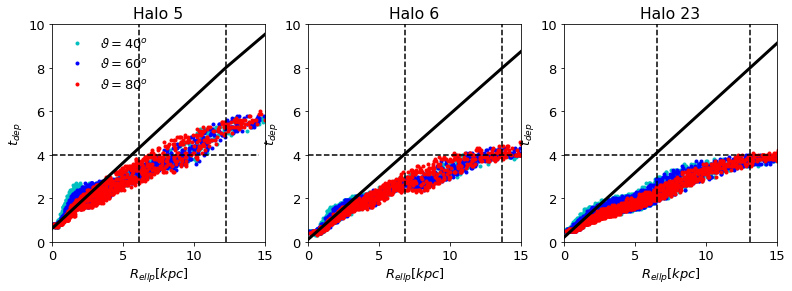}
\centering\includegraphics[width=16cm]{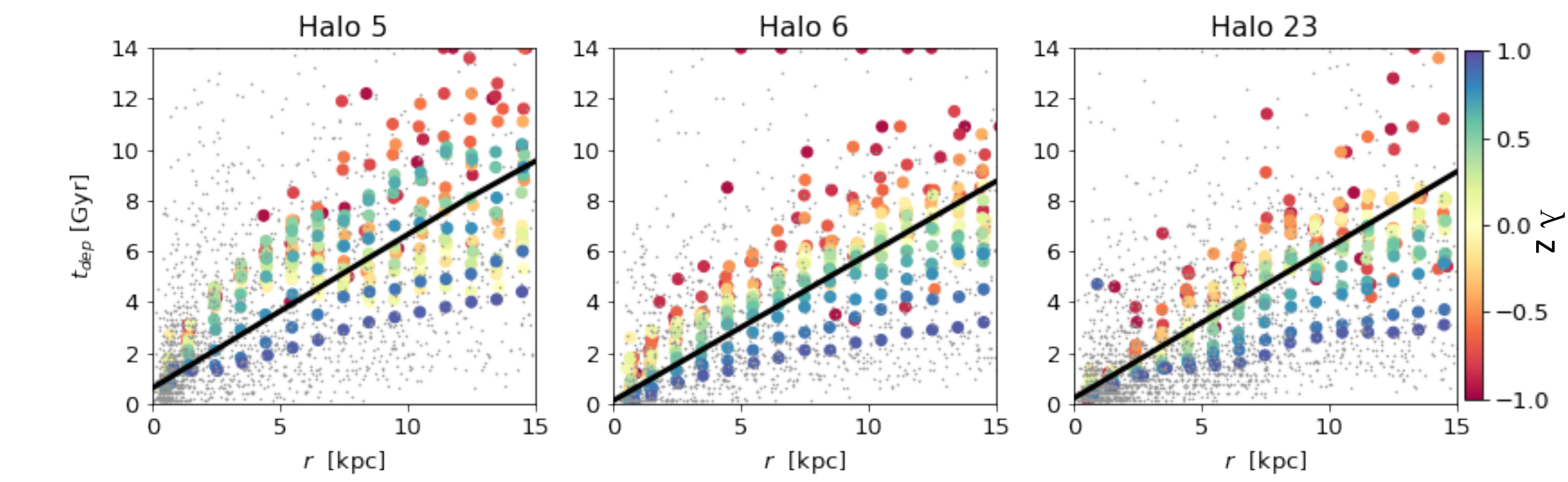}
\caption{{\bf Upper panels:} correlation of $t_{\rm dep}$ and elliptical radius $R_{\rm ellp}$ for all nine mock galaxies. $t_{\rm dep}$ at each position is derived by ($t_{\rm obs}$, $Z_{\rm obs}$) according to the theoretic relation $Z(t|t_{\rm dep})$ in Figure~\ref{fig:agemet_Ryan}. The two vertical dashed lines in each panel indicate $R_e$ and $2R_e$. The thick black lines $t_{\rm dep}(r) = a r + b$ are determined by two points: (0, $t_{\rm dep, min}$) and ($R_e$, 4 Gyr). {\bf Bottom panels:} correlation of $t_{\rm dep}$ and intrinsic radius $r$ in the three simulations. The gray dots represent particles in the simulation, the colored dots denote particles binned in the phase space $r$ vs. $\lambda_z$, colored by their circularity $\lambda_z$ as shown by the colorbar. The black lines are the same as the upper panels. Note that the y axis has different scales in the upper and bottom panels.}
\label{fig:tdepr}
\end{figure*}

From the observed age ($t_{\rm obs}$) and metallicity ($Z_{\rm obs}$) at each position, we can derived a corresponding $t_{\rm dep}$ according to the theoretical relation $Z(t|t_{\rm dep})$ as shown in Figure~\ref{fig:agemet_Ryan}. Here in Figure~\ref{fig:tdepr}, we show the correlation of $t_{\rm dep}$ with elliptical radius $R_{\rm ellp}$ across the observational plane.
$t_{\rm dep}$ is almost linearly correlated with the elliptical radius $R_{\rm ellp}$. $t_{\rm dep}$ is smaller in the inner regions with large mass density, and larger in the outer regions with small mass density. 

The observed metallicity maps have a narrow region of metallicity due to projection effects, compared to the intrinsic metallicity distribution of the particles. Thus the $t_{\rm dep}$ we derived in this way will likely underestimate the true maximum depletion time (and range of depletion times).

As shown in Figure \ref{fig:agemet_Ryan}, the observed age and metallicity distributions are bounded by depletion times which correlate with the projected radius of the bins. The upper panels of Figure \ref{fig:tdepr} shows the explicit link between the derived depletion time, and the projected elliptical radius of each bin for the 9 mock galaxy projections. In the bottom panel, we show the relation of $t_{dep}$ with the intrinsic radius $r$ for the particles in the simulations, each gray dot represent one particle in the simulation (we plot 1/1000), the colored dots denote particles binned in the phase space $r$ vs. $\lambda_z$, colored by their circularity $\lambda_z$ as shown by the colorbar.

To correct for the loss of information (primarily the suppression of the width of projected metallicity and age distributions, compared to the true particle distributions), we compute a depletion time correlation with radius which extends to larger values than the (biased) projected bins.  We find that a more complete range of depletion times (important for the most metal poor orbits) are encompassed if we fit a linear relation $t_{dep}(r) = ar + b$ to two points:  (1) $t_{dep,min}$ based on the observed age and metallicity at r=0, (2) $t_{dep}(Re) = 4 Gyr$. This relation which we adopt for this work is shown as the black line in Figure~\ref{fig:tdepr}. The relations are generally consistent with the relation of $t_{dep}$ with the intrinsic radius $r$ in the simulations.


\section{Figures for all nine galaxies}
\begin{figure*}
\centering\includegraphics[width=12cm]{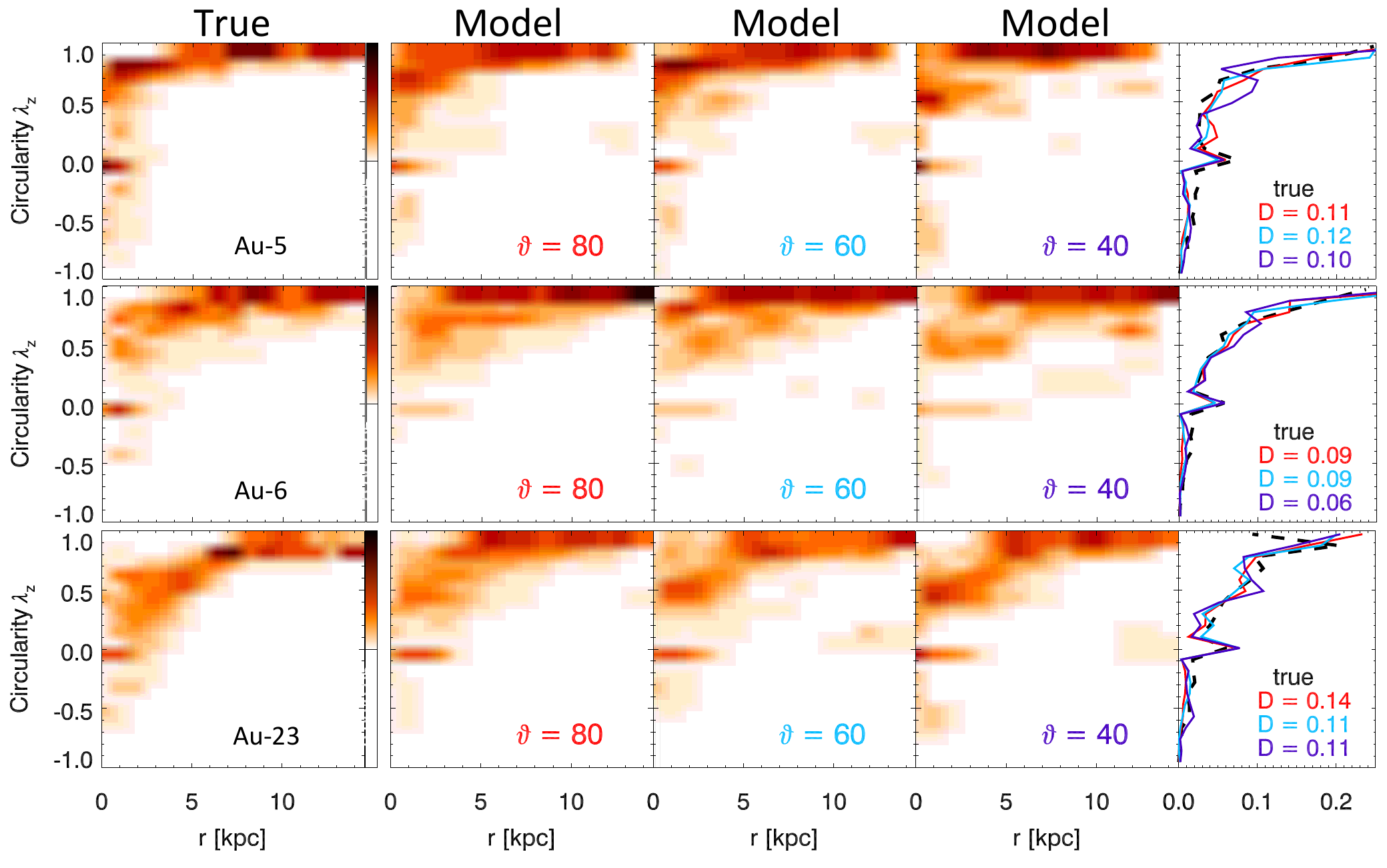}
\caption{The stellar orbit distribution in
  $\lambda_z$ vs. $r$, for Au-5, Au-6, Au-23 from top to bottom.
  The first column is the true stellar orbit
  distributions from simulations. The rest
  columns are the distribution of orbits in our
best-fitting models for mock data with $\vartheta = 80^o, 60^o, 40^o$ from
left to right. The last column is the marginalized $\lambda_z$
distribution. The black dashed curves are the true
from simulations, red, blue, purple solid curves represent those
from models for mock data with $\vartheta = 80^o, 60^o, 40^o$,
respectively. The D-statistics $D$ calculated from KS test comparing total
$\lambda_z$ distribution from our model to the true from simulations are labeled with
the corresponding colors. }
\label{fig:rlz_all}
\end{figure*}

\begin{figure*}
\centering\includegraphics[width=12cm]{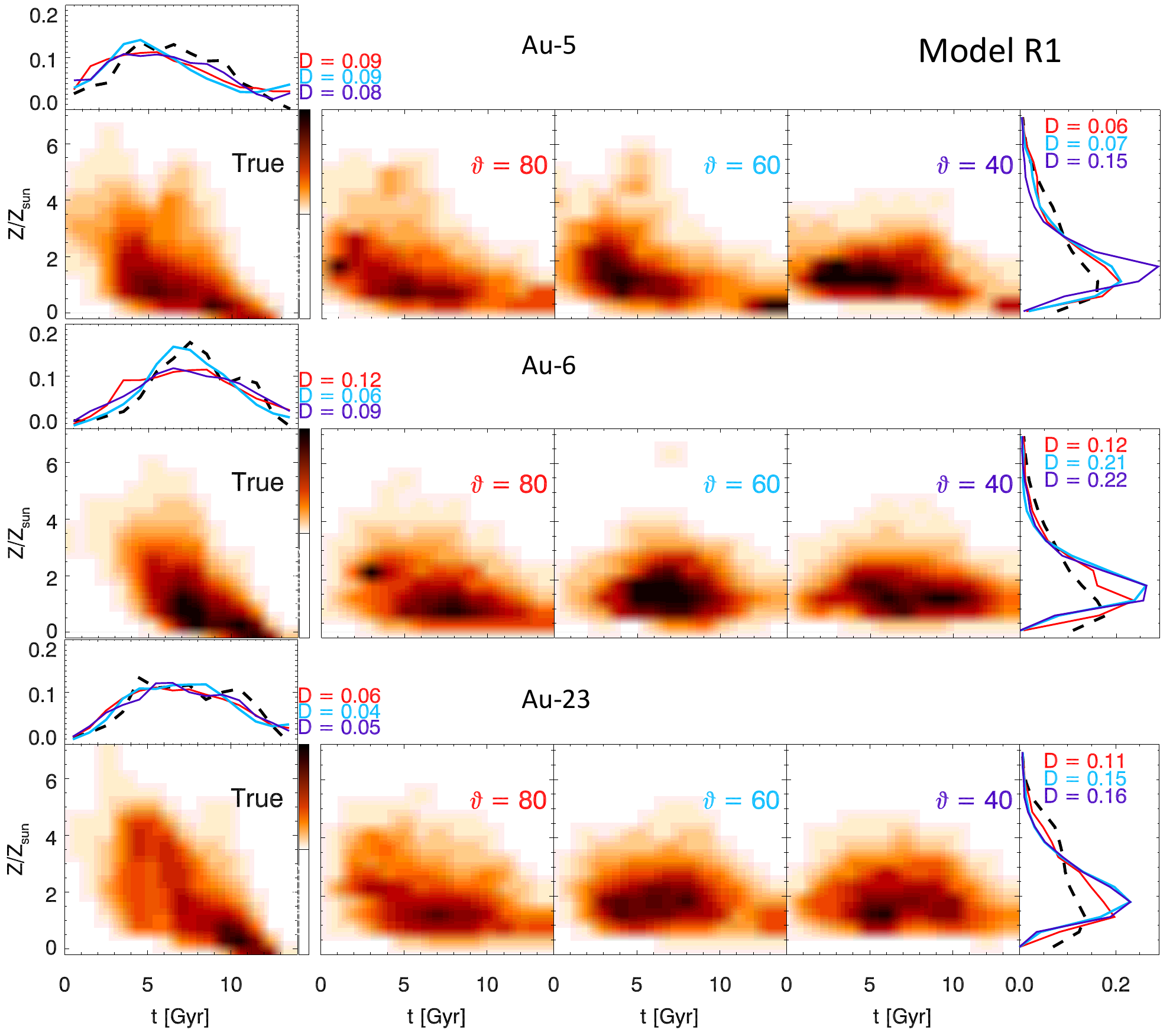}
\caption{The intrinsic age-metallicity distribution, for Au-5, Au-6,
  Au-23 from top to bottom, for model R1.
 For each halo, the panel labeled with `True'  is the true distribution on Age vs. $Z$ of particles in the
  simulation. The following panels from left to right are those obtained by our model for
  mock data with inclination angle $\vartheta$ of $\vartheta = 80^o,
  60^o, 40^o$, respectively. The
  contours are smoothed by the distribution of $t$ and $Z$ from MCMC sampling. 
  The upper subpanel is the marginalized age distribution and the
  right sub-panel is the marginalized metallicity $Z$ distribution. The black
dashed curve is the true, red, blue, purple solid curves represent
that from model for mock data with $\vartheta = 80^o, 60^o, 40^o$,
respectively, the D-statistics $D$ calculated from 1D KS test are labeled with the
corresponding colors.}
\label{fig:tZ_all_noprior}
\end{figure*}

\begin{figure*}
\centering\includegraphics[width=12cm]{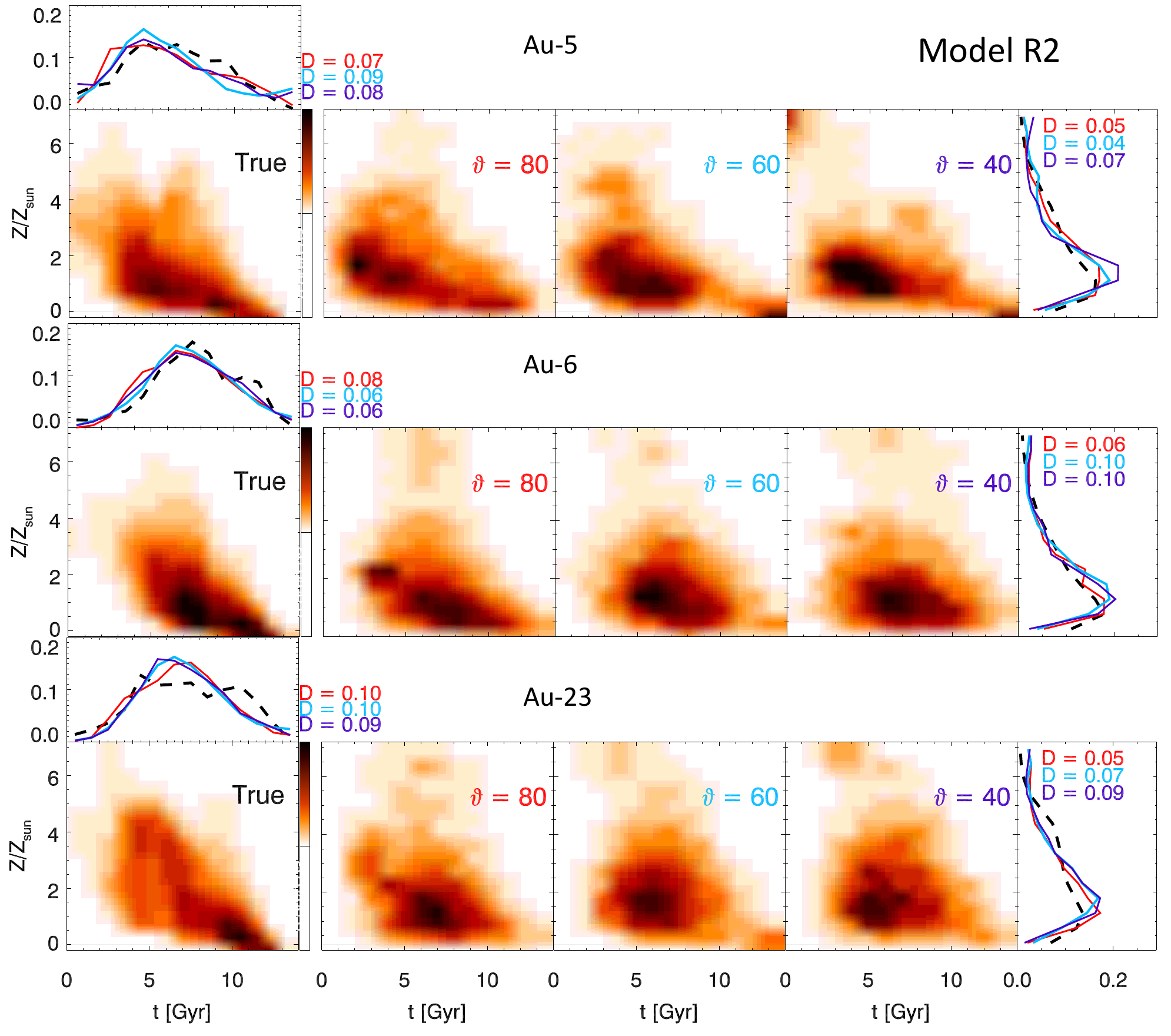}
\caption{ Similar to Figure~\ref{fig:tZ_all_noprior}, but with Model R2. The major tracks on age vs. metallicity distribution are
recovered better than model R1, the D-statistics $D$ from 1D KS test for metallicity distribution are also smaller. 
}
\label{fig:tZ_all}
\end{figure*}

\begin{figure*}
\centering\includegraphics[width=12cm]{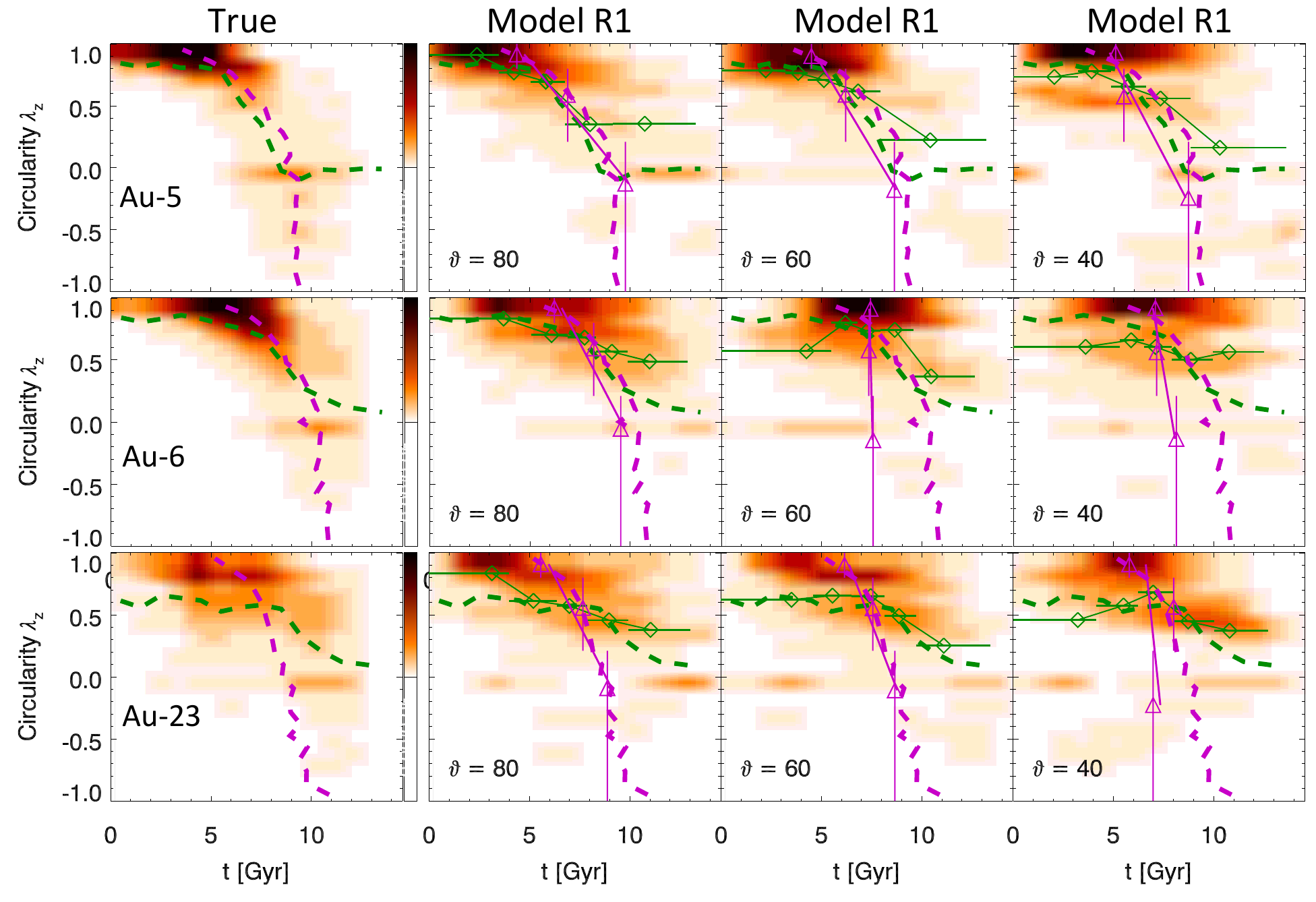}
\caption{The intrinsic correlation of age vs circularity, for Au-5,
  Au-6, Au-23 from top to bottom, for model R1.  The First column are true
  distributions of the particles from simulations, darker color
  indicates higher probability density, the magenta dashed curves are average $t$ as a
  function of $\lambda_z$, while the green dashed curves are average $\lambda_z$ as a function of age $t$ for the true distributions.
  The following panels from left
  to right are those obtained by our models for mock data with
  inclination angle $\vartheta$ of $\vartheta = 80^o, 60^o, 40^o$,
  respectively. In each panel, the magenta triangles
  are average $t$ as a function of $\lambda_z$, the magenta lines are linear fits to the triangles.
  While the green diamonds are average $\lambda_z$ as a function of age $t$ from the model. }
\label{fig:tlz_all_noprior}
\end{figure*}

\begin{figure*}
\centering\includegraphics[width=12cm]{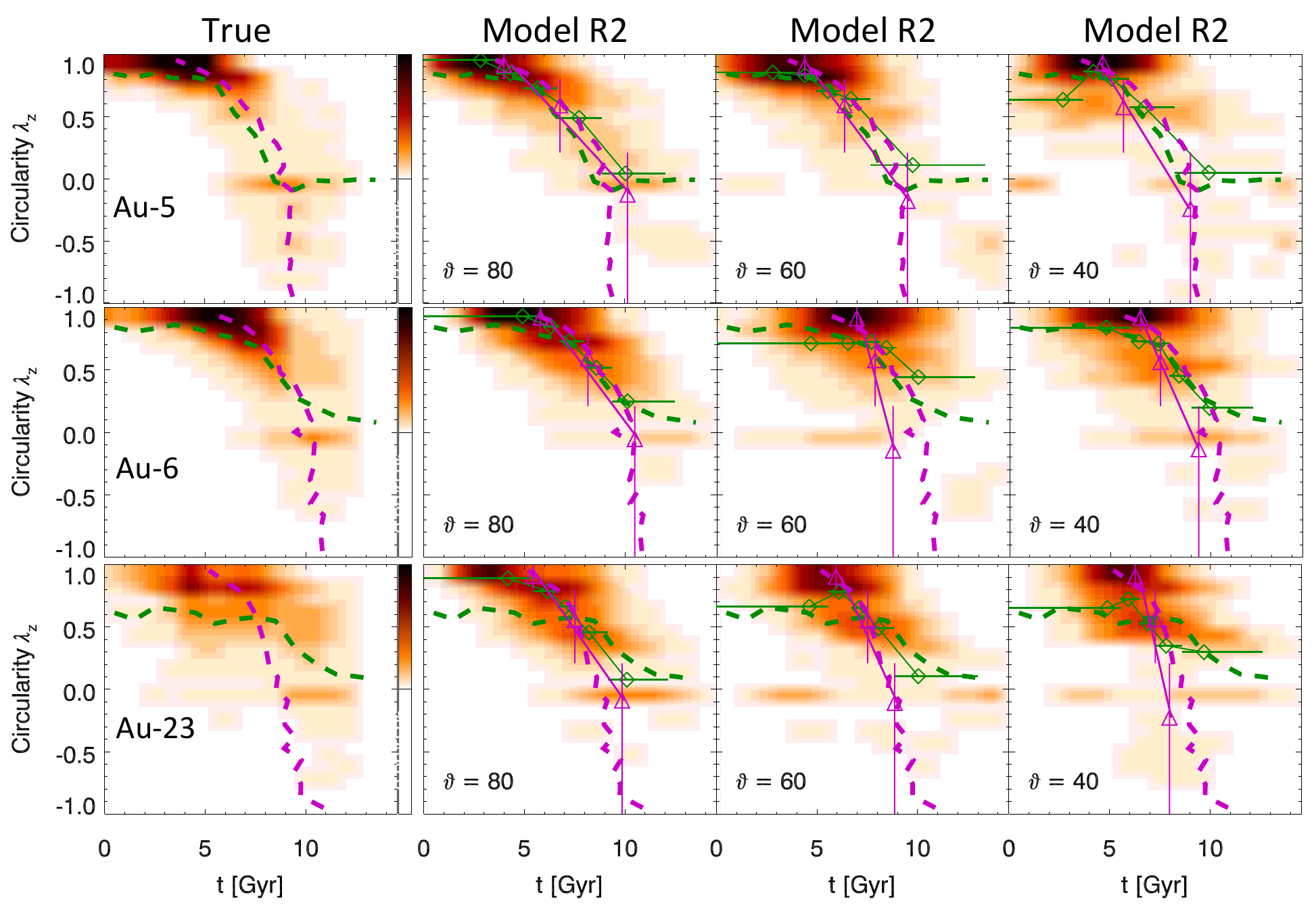}
\caption{Similar to
  figure~\ref{fig:tlz_all_noprior}, but for model R2, in which we use $t = t_0 + p\lambda_z$ (the blue line fitting the blue triangles)
  from model R1 as priors of $t_k$ in fitting the age map. Model R2 matches the age vs. $\lambda_z$ correlation in the simulation better than Model R1, especially for face-on galaxies.}
\label{fig:tlz_all}
\end{figure*}

Similar to figures we show for the galaxy Au-6 $\vartheta=80^o$ in Section~\ref{S:result}. Figure~\ref{fig:rlz_all} shows the stellar orbit distribution on $r$ vs. $\lambda_z$ comparing with the true from simulation and those from our models for all nine galaxies. Figure~\ref{fig:tZ_all_noprior} and Figure~\ref{fig:tZ_all}
show the stellar population distribution $t$ vs. $Z$ from our models for all nine galaxies, with different priors of R1 and R2, respectively. Figure~\ref{fig:tlz_all_noprior} and Figure~\ref{fig:tlz_all} are the correlation of age $t$ and circularity $\lambda_z$ for all nine galaxies, with different priors of R1 and R2, respectively.

\begin{figure*}
  \centering\includegraphics[width=17cm]{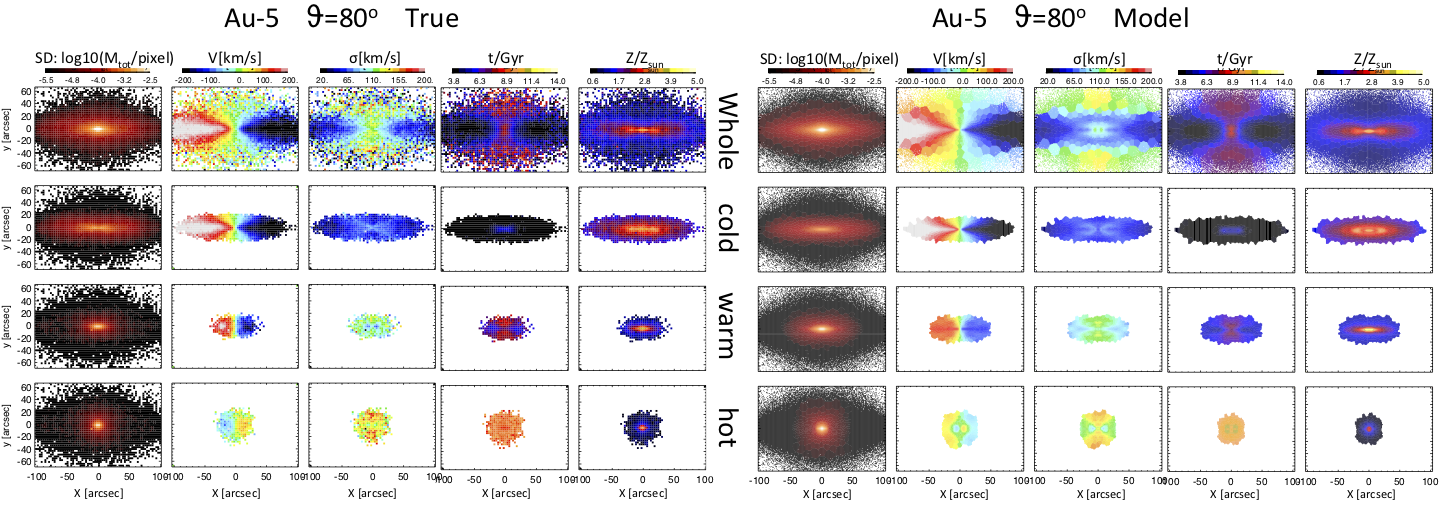}
  \centering\includegraphics[width=17cm]{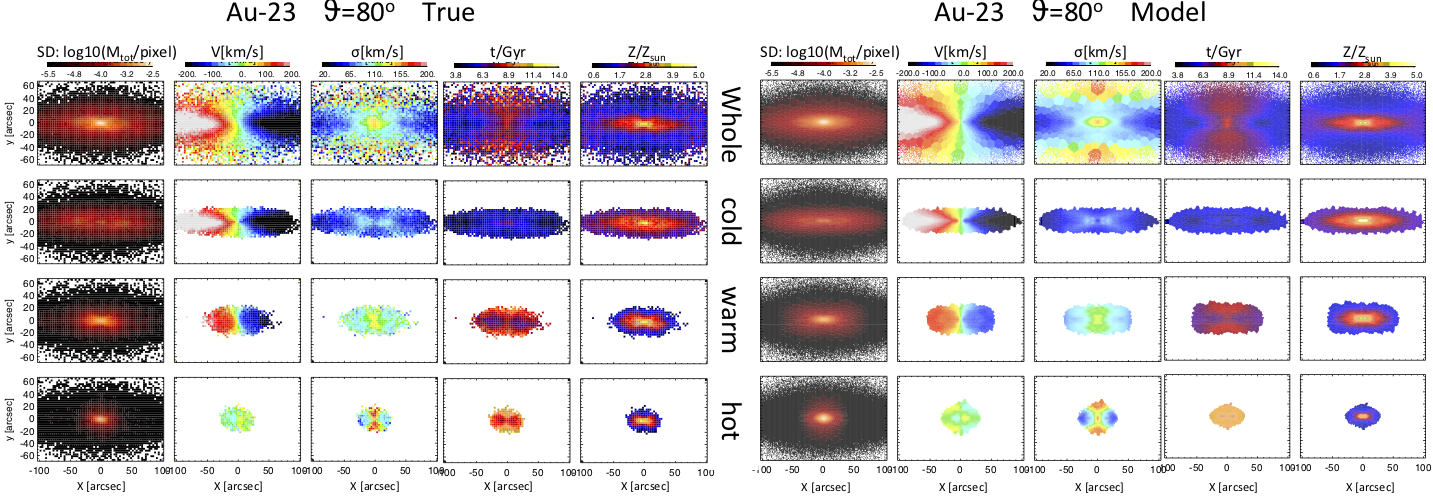}
\caption{Surface mass density/brightness, mean velocity, velocity dispersion, age, and metallicity maps of the the
  whole galaxy,  cold, warm, and hot + CR component (from top tp
  bottom) of Au-5 $\vartheta =
  80^o$ and Au-23 $\vartheta =
  80^o$. The left panels are the
  true values from the simulation, the right panels are rebuilt
  by orbital bundles from our model R2. These galaxies are at a distance of 30 Mpc, $1'' = 145 \, {\rm pc}$.}
\label{fig:halo523_ztmaps_kin3}
\end{figure*}

\begin{figure}
  \centering\includegraphics[width=7cm]{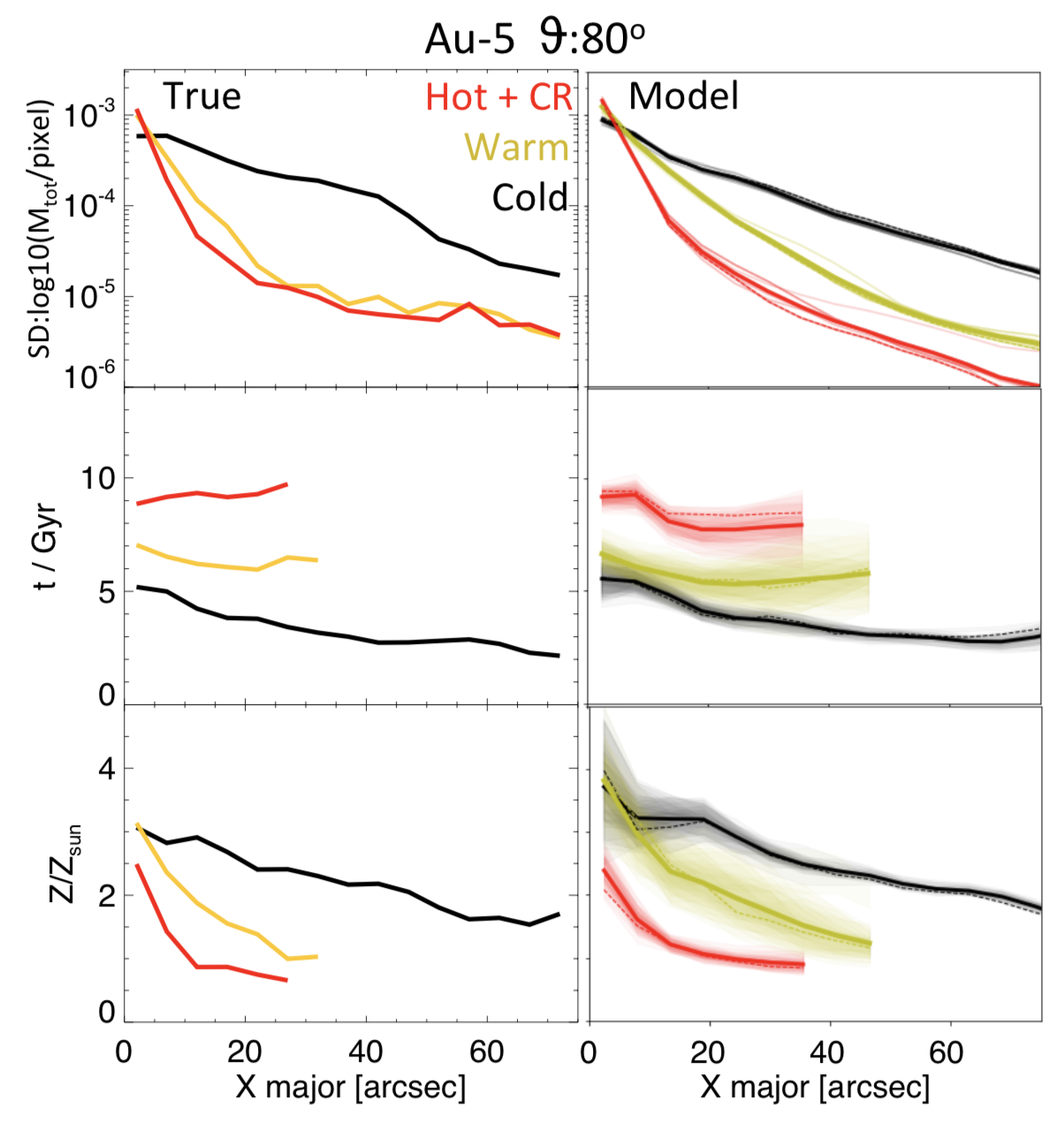}
  \centering\includegraphics[width=7cm]{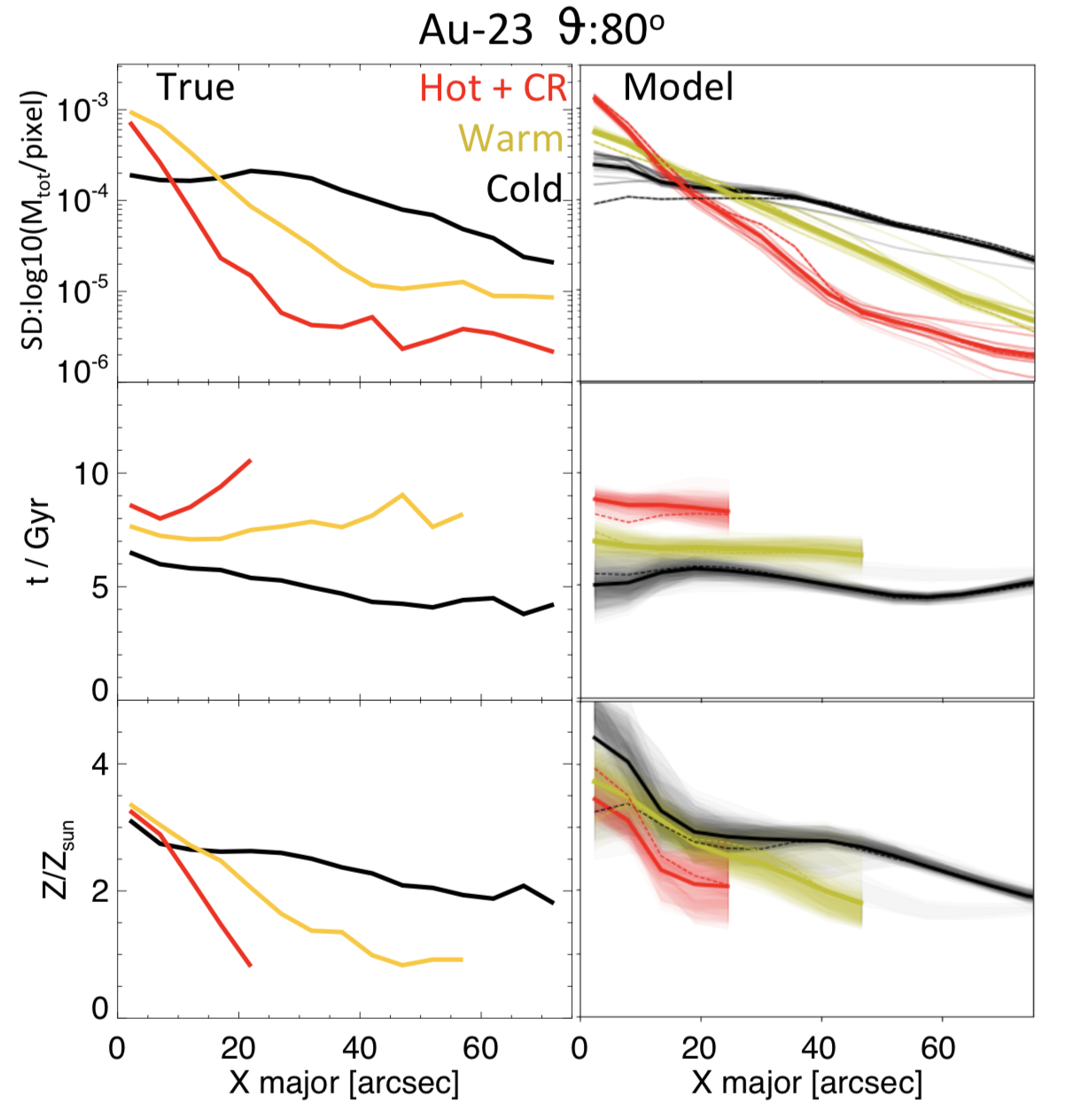} 
\caption{Surface mass density/brightness profile, age, and metallicity along major
  axis ($1'' = 145 \, {\rm pc}$) of the cold, warm, and hot component, comparison between
  true and those built by model, similar to
  Figure~\ref{fig:halo6_zt_kin3} for Au-6 $\vartheta = 80^o$. We generally recover the surface brightness, age and metallicity profiles of each component well. Except for Au-23, we over-estimate metallicity of cold component in the inner regions, thus resulting in a stronger negative metallicity gradient of this component than the true. }
\label{fig:halo523_zt_kin3}
\end{figure}

Similar to figures we show fro Au-6 $\vartheta=80^o$ in
Section~\ref{S:decomp}.
Figure~\ref{fig:halo523_ztmaps_kin3} are surface brightness, mean velocity, velocity dispersion, age and
metallicity maps of cold, warm, hot+CR components, comparing with true
from simulation with our model R2, for Au-5 $\vartheta = 80^o$
and Au-23 $\vartheta=80^o$. 
Figure~\ref{fig:halo523_zt_kin3} are the age, metallicity profiles along major axis.

\end{document}